\documentclass{article}
\usepackage{graphicx} % Required for inserting images

\usepackage[hidelinks]{hyperref} 
\usepackage{bbding}
\usepackage{amsthm}
\usepackage{shuffle}
\usepackage{mathtools}
\usepackage{tensor}
\usepackage{mdframed}
\usepackage{graphicx}
\usepackage{caption}
\usepackage{subcaption} % or use \usepackage{subfig}
\usepackage{float}
\usepackage{amsfonts}
\usepackage{tikz}
\usepackage{simplewick}
\usepackage{circuitikz}
\usepackage{xcolor}
\usetikzlibrary{ decorations.markings}
\usetikzlibrary{cd}
\usepackage{eucal}
\usepackage{physics}
\usepackage[english]{babel}
\usepackage{enumerate}

\theoremstyle{plain}
\newtheorem{theorem}{Theorem}[section]
\newtheorem{lemma}[theorem]{Lemma}

\newtheorem{corollary}[theorem]{Corollary}
\newtheorem{proposition}[theorem]{Proposition}

\theoremstyle{definition}
\newtheorem{definition}[theorem]{Definition}
\newtheorem{example}[theorem]{Example}

\theoremstyle{remark}
\newtheorem*{remark}{Remark}

\newcommand{\calC}{{\mathcal{C}}}

\newcommand{\RR}{\mathbb{R}}
\newcommand{\CC}{\mathbb{C}}
\newcommand{\ZZ}{\mathbb{Z}}
\newcommand{\NN}{\mathbb{N}}

\usepackage[english]{babel}

\newcommand{\Tcal}{\mathcal{T}}
\newcommand{\dT}{\cdot_{{}^\Tcal}}
\newcommand{\starT}{\mathbin{\star_{\scriptscriptstyle \Tcal}}}
\DeclareMathOperator{\Aut}{Aut}
\newcommand{\starH}{\mathbin{\star_{\scriptscriptstyle H}}}
\newcommand{\Gcal}{\mathcal{G}}
\DeclareMathOperator{\supp}{supp}
\newcommand{\QH}{{\mathcal Q}_H}
\newcommand{\QT}{{\mathcal Q}_\Tcal}
\newcommand{\QTbar}{{\mathcal Q}_{\overline{\Tcal}}}
\newcommand{\hatphi}{\hat\phi}

\newcommand{\source}{\mathsf{s}}
\newcommand{\target}{\mathsf{t}}
\DeclareMathOperator{\inv}{inv}
\newcommand{\IN}{\mathrm{in}}
\newcommand{\out}{\mathrm{out}}
\DeclareMathOperator{\VP}{VP}
\newcommand{\ith}{\text{th}}
\newcommand{\Op}{\mathrm{Op}}

\usepackage[a4paper,top=2cm,bottom=3cm,left=3cm,right=3cm,marginparwidth=1.75cm]{geometry}

\usepackage{amsthm}
\usepackage{amssymb}
\usepackage{mathrsfs}
\usepackage{siunitx}
\PassOptionsToPackage{hyphens}{url}
\usepackage{hyperref}
\usepackage{cleveref}
\usepackage[utf8]{inputenc}
\usepackage[right]{lineno}
\usepackage{csquotes}
\usepackage{booktabs}
\usepackage{longtable}
\usepackage{adjustbox}
\usepackage{array}
\usepackage{url}
\usepackage{titlesec}
\usepackage{authblk}
\usepackage{xcolor} % Load the xcolor package for color options

\usepackage[colorinlistoftodos,textsize=footnotesize]{todonotes}

\titleformat{\subsection}
  {\mdseries\itshape\large} % Medium series, italic shape, and large font size
  {\thesubsection}{1em}{} % Numbering, spacing, and the section title itself

\usepackage[english]{babel}
\usepackage[style=authoryear,backend=biber,natbib=true,maxcitenames=2,uniquelist=false]{biblatex}
\DeclareNameAlias{sortname}{family-given}
\DeclareNameAlias{default}{family-given}

\renewbibmacro{in:}{}
\DeclareFieldFormat[article]{title}{\mkbibquote{#1}\addcomma}
\DeclareFieldFormat[book]{title}{\mkbibemph{#1}\addcomma}
\DeclareFieldFormat[bookinbook]{title}{\mkbibemph{#1}\addcomma}
\DeclareFieldFormat[inbook]{title}{\mkbibquote{#1}\addcomma}
\DeclareFieldFormat[incollection]{title}{\mkbibquote{#1}\addcomma}
\DeclareFieldFormat[inproceedings]{title}{\mkbibquote{#1}\addcomma}
\DeclareFieldFormat[manual]{title}{\mkbibemph{#1}\addcomma}
\DeclareFieldFormat[misc]{title}{\mkbibemph{#1}\addcomma}
\DeclareFieldFormat[thesis]{title}{\mkbibemph{#1}\addcomma}
\DeclareFieldFormat[unpublished]{title}{\mkbibquote{#1}\addcomma}
\DeclareFieldFormat[patent]{title}{\mkbibemph{#1}\addcomma}
\DeclareFieldFormat[report]{title}{\mkbibemph{#1}\addcomma}
\DeclareFieldFormat[online]{title}{\mkbibquote{#1}\addcomma}
\DeclareFieldFormat[software]{title}{\mkbibemph{#1}\addcomma}
\DeclareFieldFormat[booklet]{title}{\mkbibemph{#1}\addcomma}
\DeclareFieldFormat[periodical]{title}{\mkbibemph{#1}\addcomma}
\DeclareFieldFormat[standard]{title}{\mkbibemph{#1}\addcomma}

\DeclareFieldFormat[article]{journaltitle}{\iffieldundef{shortjournal}{\mkbibemph{#1}\addcomma}{\mkbibemph{\printfield{shortjournal}}\addcomma}}
\DeclareFieldFormat{volume}{\bibstring{volume}~#1}
\DeclareFieldFormat{number}{\bibstring{number}~#1}

\DefineBibliographyStrings{english}{
  volume = {Vol.},
  number = {No.}
}

\renewbibmacro*{volume+number+eid}{%
  \printfield{volume}%
  \setunit*{\addspace}%
  \printfield{number}%
  \setunit{\addcomma\space}%
  \printfield{eid}}

\renewbibmacro*{journal+issuetitle}{%
  \usebibmacro{journal}%
  \setunit*{\addcomma\space}%
  \usebibmacro{volume+number+eid}%
  \setunit{\addcomma\space}%
  \usebibmacro{issue+date}}

\renewbibmacro*{publisher+location+date}{%
  \printlist{publisher}%
  \iflistundef{location}
    {\setunit*{\addcomma\space}}
    {\setunit*{\addcolon\space}}%
  \printlist{location}%
  \setunit*{\addcomma\space}%
  \usebibmacro{date}}

\DeclareCiteCommand{\cite}[\mkbibparens]
  {\usebibmacro{prenote}}
  {\usebibmacro{citeindex}%
   \usebibmacro{cite}}
  {\multicitedelim}
  {\usebibmacro{postnote}}

\renewbibmacro*{cite:labelyear+extrayear}{%
  \iffieldundef{labelyear}
    {}
    {\printtext[bibhyperref]{%
       \printfield{labelyear}%
       \printfield{extrayear}}}}

\renewbibmacro*{cite:labeldate+extradate}{%
  \iffieldundef{labelyear}
    {}
    {\printtext[bibhyperref]{%
       \printfield{labelyear}%
       \printfield{extradate}}}}

\AtEveryBibitem{
  \clearfield{month}
  \clearfield{day}
  \ifentrytype{book}{
    \clearlist{location}
  }{}
}

\DefineBibliographyStrings{english}{
  andothers = {\textit{et al.},}
}

\DeclareFieldFormat[article]{volume}{\bibstring{jourvol}\addnbspace #1}
\DeclareFieldFormat[article]{number}{\bibstring{number}\addnbspace #1}
\DeclareFieldFormat[article]{volume}{Vol. #1}
\DeclareFieldFormat[article]{number}{No. #1}
\DeclareFieldFormat{url}{\bibstring{available at}\addcolon\space\url{#1}}
\DeclareFieldFormat{urldate}{\mkbibparens{accessed \addspace#1}}

\DeclareFieldFormat{urldate}{%
  \mkbibparens{accessed\space%
    \thefield{urlday}\addspace%
    \mkbibmonth{\thefield{urlmonth}}\addspace%
    \thefield{urlyear}}}

\crefformat{figure}{#2Figure~#1#3}
\Crefformat{figure}{#2Figure~#1#3}
\crefformat{table}{#2Table~#1#3}
\Crefformat{table}{#2Table~#1#3}
\crefformat{section}{#2Section~#1#3}
\Crefformat{section}{#2Section~#1#3}

\title{Causal set QFT from deformation quantisation}
\author{Eli Hawkins$^{a}$, Kasia Rejzner$^{a}$ and Stav Zalel$^{b,c}$}

\affil{$^{a}$Department of Mathematics, University of York,
Heslington, York YO10 5DD, United Kingdom\\
$^{b}$Department of Applied Mathematics and Theoretical Physics,
University of Cambridge, CB3 0WA, United Kingdom \\
$^{c}$Homerton College, University of Cambridge, CB2 8PQ, United Kingdom}

\begin{document}

\maketitle

\begin{abstract}
	 We explicitly show the equivalence between three approaches to constructing QFT expectation values on a causal set: the operator approach, the path-integral approach and the algebraic approach (using deformation quantisation). In the algebraic approach, observables are modelled as polynomial functionals of off-shell field configurations, and quantisation maps translate star products of functionals into products of field operators; in-in correlators are computed using M\o ller operators, while in-out correlators and matrix elements between particle states use the time-ordering operator applied to the formal $S$-matrix. Evaluating M\o ller operators on a causal set is tractable, and we give examples of how it can be done in practice. We compare diagrammatic computations of M\o ller operators and expectation values of field operators, emphasizing the equivalence of the approaches. We give a double path integral expression for the vacuum expectation value of a M\o ller operator and, more generally, for in-in correlators of anti-causally and causally ordered products of interacting fields. Unlike in the continuum, both sides of these identities are independently well defined.
\end{abstract}

\tableofcontents

\section{Introduction}

Understanding the interplay between matter and spacetime near the Planck scale is a challenge shared by all approaches to quantum gravity. In causal set theory, where spacetime takes the form of a causal set---a discrete and dynamical Lorentzian lattice---one is therefore motivated to seek a framework for quantum field theory on a causal set background. The framework developed to date allows one to define and make computations in interacting scalar theories on a fixed causal set, in analogy to the continuum's quantum field theory on a curved spacetime with no dynamical gravity or back-reaction.

This paper brings together two different approaches to the problem of quantum field theory on causal sets, namely: perturbative algebraic quantum field theory and the operator/path integral approach. The two approaches emphasise different structures. The algebraic approach is covariant and separates the construction of the algebra of observables from the choice of a state. The operator approach works in a fixed Fock representation and connects directly to the histories (path integral) formulation of quantum theory and to scattering. Relating them makes the tools of each approach available in the other.

Our main tool is deformation quantisation \cite{Bayen:1977ha,Bayen:1978II}. It describes a quantum system in terms of its classical observables: the commutative product of functions is deformed into a non-commutative, associative \emph{star product}, which encodes the products of operators in the quantum theory. A \emph{quantisation map} sends classical observables to operators and intertwines a star product with the operator product. Different quantisation maps correspond to different operator orderings, such as symmetric (Weyl) ordering or normal ordering, and the associated star products are related by ``gauge'' transformations. General constructions of star products on finite-dimensional symplectic and Poisson manifolds were given in \cite{Fedosov:1994} and \cite{Kontsevich:2003}, respectively. The first applications of deformation quantisation to quantum field theory are \cite{Dito:1990,Dito:1993}, where the normal star product was used for the free scalar field. Formal deformation quantisation is also one of the ingredients of perturbative algebraic quantum field theory (pAQFT) \cite{Dutsch:2000nh,Dutsch:2001dq,Brunetti:2009qc,Fredenhagen:2012sb}; see \cite{Rejzner:2016,Dutsch:2019} for reviews. In pAQFT, the classical observables are functionals of field configurations. In this paper, $\phi_x$ denotes the evaluation functional $\phi\mapsto\phi(x)$, which assigns to a field configuration its value at the point $x$ of the causal set. It is a classical observable, a coordinate on the space of configurations. The field operator $\hatphi_x$ is its image under a quantisation map, and polynomials in the $\phi_x$ are mapped to suitably ordered products of the $\hatphi_x$.

Free and interacting algebraic quantum field theory on causal sets was put forward in \cite{Dable-Heath:2019sej}. The algebraic approach is particularly well-suited to causal set theory, thanks to its intrinsically covariant nature. It replaces the canonical formulation of operators and Hilbert spaces with algebras of  observables  associated with causal regions, and the state of the system is a map from the algebra of observables to complex numbers---these notions are readily adaptable to the causal set setting. One way to construct these algebras is by means of (formal) deformation quantisation: observables are modeled as functionals and the pointwise product on functionals is deformed to a star product that embodies the non-commutativity of operators in the canonical approach. Interactions are introduced via the formal $S$ matrix and M\o ller operators.
In \cite{Hawkins:2016ddq}, a diagrammatic expansion of star products and M\o ller operators was derived, yielding the Feynman diagrams of this approach.

Throughout, we work off-shell: functionals are defined on all field configurations $\phi\in\RR^\calC$, and no field equation is imposed. This is the approach of \cite{Hawkins:2016ddq}. In the alternative on-shell approach, the field equation is imposed before quantisation, and Fedosov's construction has been extended to that setting \cite{Collini:2016}. Working off-shell is natural on a causal set, where the free theory is specified by its two-point function rather than by a field equation. It is also what makes the constructions in this paper well defined. The time-ordered product of functionals and the corresponding quantisation map exist off-shell, whereas causal ordering is not a well-defined map on the algebra of field operators, just as time-ordered products of on-shell fields in the continuum are not (see Section~\ref{sec:op}). The histories formulation is off-shell as well, since one integrates over all configurations.

The operator approach \`{a} la Sorkin and Johnston was first proposed in \cite{Johnston:2008za,Johnston:2009fr} (see also \cite{Sorkin:2017fcp}) and its path integral description was derived in \cite{Sorkin:2011pn}. This last paper concluded with a concrete proposal for introducing interactions via the path integral. This proposal was realised verbatim in \cite{Albertini:2024srq}. The Sorkin-Johnston operator formalism was also extended in the same work to include interactions and was applied in deriving the corresponding Feynman diagrams. A definition of scattering amplitudes on a causal set was also proposed there, and a reduction formula for matrix elements between particle states was derived in \cite{Zalel:2026srq}.

Here, we bring these strands together by showing explicitly how the quantities defined in the operator/path integral approach---in-in expectation values, in-out correlation functions and matrix elements defined by particle states---are computed in the language of algebraic quantum field theory and deformation quantisation.

The key tool is the quantisation map $\QH$, which is normal ordering with respect to the Fock vacuum defined by the two-point function $W$ (Lemma~\ref{Expectation lemma}). It intertwines the Wick star product $\starH$ with the operator product, and the vacuum expectation value of $\QH(F)$ is simply $F(0)$. We choose normal ordering because it is adapted to the operator approach, in which all quantities are vacuum expectation values or matrix elements in the Fock representation. 
	
Causal ordering corresponds to the quantisation map $\QT=\QH\circ\Tcal$. Unlike in the continuum, we do not normal order the interaction. The coincidence limits $W_{xx}=\Delta^F_{xx}$ are finite, so $V(\hatphi_x)$ is a well-defined operator and no renormalisation is needed. Consequently, the diagrams of Section~\ref{sec: diagramatic} include loops at internal vertices. These are absent in \cite[Prop.~6.42]{Hawkins:2016ddq}, where $\Tcal$ acts as the identity on local functionals, that is, the interaction is normal ordered.

This paper continues as follows. Section 2 provides a brief technical
introduction. In Section 3 we present a series of propositions (Propositions~\ref{proposition: free}--\ref{prop: matrix element}) that act as a dictionary, translating between the two formalisms. For every polynomial $F$, the causally ordered in-in correlator is ${}_\IN\langle0|C[F(\hatphi^H)]|0\rangle_\IN=R_{H,gV}(\Tcal F)(0)$, and the in-out correlator is ${}_\out\langle0|C[F(\hatphi^H)]|0\rangle_\IN=\Tcal(\mathscr SF)(0)$. We also prove the reduction formula for particle states directly from the double path integral (Proposition~\ref{prop: matrix element}). This is the algebraic analog of the result of \cite{Zalel:2026srq}. In Section~\ref{sec: examples} we give examples of how various relevant quantities can be computed in practice using the algebraic approach. In Section~\ref{sec: diagramatic} we compare the diagrammatic approaches of \cite{Hawkins:2016ddq} and \cite{Albertini:2024srq}, respectively, demonstrating again their equivalence.
In Section~\ref{sec: path integral} we prove a theorem, giving double path integral expressions for the product $\overline{\Tcal}(G)\starH\mathcal{T}(F)$ and the M\o ller operator $R_{H, gV}(\mathcal{T}F)$, evaluated at zero. Unlike in the continuum, where the analogous expressions are formal, both sides of our expressions are well-defined independently and can be computed. As a consequence, we obtain the double path integral formula for the in-in correlators $\langle\overline C[G(\hatphi^H)]\,C[F(\hatphi^H)]\rangle$ for arbitrary $G$ directly in the algebraic approach. In the operator approach this has been done in \cite{albertini:2021,Jubb:2023mlv,Albertini:2024srq}. Section~\ref{sec: conclusions} contains our conclusions, and Appendix~\ref{appendix: sec diagrams} the derivation of the operator diagrammatic expansion.

\section{Technical background}
 We will be concerned with real scalar QFTs on a finite causal set, $\calC$ with causal order $\prec$. Write $N=|\calC|$ and fix a natural labelling $\calC=\{1,\ldots,N\}$, so that $x\prec y$ implies $x<y$. We take $\hbar>0$. All interaction expansions and their identities are understood coefficientwise in $g$, with polynomial coefficients; empty products are $1$.

\subsection{Operator formalism}\label{sec:op}
In the approach detailed in \cite{Sorkin:2017fcp}, the Wightman (two-point) function furnishes the starting point for constructing a QFT in a Gaussian state. On a causal set, the Wightman function takes the form of a complex $N\times N$ matrix, $W_{xy}$, that is Hermitian, positive and causal (\textit{i.e.} $W_{xy}\in\mathbb{R}$ whenever $x$ and $y$ are spacelike). Due to the positivity of $W$, one obtains a Fock space and field operators acting thereon by diagonalizing $W$ as,
\begin{equation}\label{W decomposition}
    W_{xy}=\sum_{\lambda}\lambda w^{\lambda}_x\overline{w^{\lambda}_y},
\end{equation}
where the sum is over a chosen orthonormal basis of positive-eigenvalue modes of $W$ and bar denotes complex conjugation. The mode label $\lambda$ also denotes its eigenvalue; modes in a degenerate eigenspace have distinct labels, and $\delta_{\lambda\mu}$ compares these labels,
and introducing the corresponding ladder operators, $a_{\lambda}$ and $a^{\dagger}_{\lambda}$, satisfying the relation,
\begin{equation}
    [a_{\lambda},a_{\mu}^{\dagger}]= \delta_{\lambda\mu},
\end{equation}
and the corresponding Fock vacuum $|0\rangle$,
\begin{equation}
    a_{\lambda}|0\rangle=0 \ \forall \ \lambda.
\end{equation}
The full Hilbert space of the theory is spanned by acting on the vacuum with all
possible combinations of the creation operators to obtain ``particle'' states of the form,
\begin{align}
    |\lambda_1\ldots \lambda_k\rangle=\frac{1}{\sqrt{\lambda_1\ldots \lambda_k}}a_{\lambda_1}^{\dagger}\ldots a_{\lambda_k}^{\dagger}|0\rangle.
\end{align}  These states use an eigenvalue-dependent normalisation: $\langle\lambda|\mu\rangle=\delta_{\lambda\mu}/\lambda$, and repeated mode occupations give the usual factorials for Fock space inner products. The eigenvectors supply the mode functions and the eigenvalues are treated like frequencies.
Finally, for each $x\in\calC$, the field operator $\hatphi_x$ acting on this Hilbert space is defined as,
\begin{equation}\label{Field definition}
    \hatphi_x=\sum_{\lambda} \sqrt{\hbar\lambda}\left(w^{\lambda}_xa_{\lambda}+\overline{w^{\lambda}_x} a_{\lambda}^{\dagger}\right).
\end{equation}
One can verify that,
\begin{equation}
    \langle 0 | \hatphi_x\hatphi_y | 0 \rangle = \hbar W_{xy},
\end{equation}
and
that the field operators satisfy the Peierls bracket commutation relation,
\begin{equation}\label{peierls bracket}
    [\hatphi_x,\hatphi_y]=i\hbar\Delta_{xy},
\end{equation}
with $i\Delta=W-\overline{W}$.
\begin{remark}
    To obtain the Sorkin-Johnston vacuum, one further imposes the ground-state condition $W\overline W=0$. We do not assume this prescription in this work. 
\end{remark}

The Feynman propagator $\Delta^F$ is an $N\times N$ matrix defined as,
\begin{equation}
    \Delta^F_{xy}=
    \begin{cases}
     W_{xy} & \text{if $x\succ y$}\\
       W_{yx} & \text{otherwise},\\
    \end{cases}\end{equation}
where $W_{xy}=W_{yx}$ if $x$ and $y$ are spacelike separated. In other words, $\hbar \Delta^F_{xy}$ is the causally-ordered 2-point function, where causal ordering (our analogue of the continuum's time ordering) orders products of field operators such that if $x\succ y$ then $\hatphi_x$ is to the left of $\hatphi_y$. We write $C[\hatphi_x\hatphi_y]$ to denote the causal ordering of $\hatphi_x\hatphi_y$. Similarly, anti-causal ordering orders products of field operators such that if $x\succ y$ then $\hatphi_x$ is to the right of $\hatphi_y$. We write $\overline{C}[\hatphi_x\hatphi_y]$ to denote the anti-causal ordering of $\hatphi_x\hatphi_y$.

\begin{remark}
Throughout, $C[\cdots]$ and $\overline C[\cdots]$ act on products of operators each of which is localised at a point of $\calC$, such as $\hatphi_x$, $F(\hatphi_x)$, or the Heisenberg fields $\hatphi^H_x$ introduced below, by reordering the factors. The result does not depend on how factors at unrelated points are ordered among themselves, since such factors commute, i.e. $[\hatphi_x,\hatphi_y]=i\hbar\Delta_{xy}=0$ if $x$ and $y$ are spacelike. For Heisenberg fields see Lemma~\ref{lemma: causal factorisation}(iii). Note that $C$ acts on the list of localised factors, not on the operator obtained by multiplying them, so it is not a well-defined map on operators. For example, if $x\prec y$, then $\hatphi_x\hatphi_y$ and $\hatphi_y\hatphi_x+i\hbar\Delta_{xy}$ are the same operator, but $C[\hatphi_x\hatphi_y]=\hatphi_y\hatphi_x$, whereas $C[\hatphi_y\hatphi_x]+i\hbar\Delta_{xy}=\hatphi_y\hatphi_x+i\hbar\Delta_{xy}$. Time-ordered products of on-shell fields in the continuum behave in the same way, whereas time-ordered products of off-shell fields are well defined. Here the off-shell counterpart is the map $\QT$ on functionals of Section~\ref{Observables}; see also the Remark after Definition~\ref{definition QT}.
\end{remark}

Let $H=\Re W$. We use the strictly retarded split $\Delta^R_{xy}=\Delta_{xy}$ for $x\succ y$ and $0$ otherwise, and put $\Delta^A=(\Delta^R)^{\mathsf T}$. Then $\Delta=\Delta^R-\Delta^A$, and
 \begin{equation}\label{eq: W and DeltaF}
     \begin{split}
          &W=\frac{i}{2}(\Delta^R-\Delta^A)+H,\\
          &\Delta^F=\frac{i}{2}(\Delta^R+\Delta^A)+H.
     \end{split}
 \end{equation}
In particular $\Delta^R_{xx}=0$ and $W_{xx}=\Delta^F_{xx}=H_{xx}$ for all $x\in\calC$, so no diagonal contact term is included in these propagators. Moreover $H\geq0$: for $\xi\in\RR^N$, $\xi^{\mathsf T}H\xi=\xi^{\mathsf T}W\xi\geq0$, since $\Im W$ is antisymmetric.

On top of this free construction, we now add local polynomial interactions in the field. Given a local interaction $gV$, where $V(\phi)=\sum_y V_y(\phi_y)$ is real-valued and $g$ is real, the evolution operator $U_{x}$ is given by the causally ordered product,
\begin{equation}\label{eq: U operator}
     \hat{U}_{x}=C\bigg[\prod_{ y\prec x} e^{-\frac{ig}{\hbar}V_y(\hatphi_y)}\bigg],
\end{equation}
where it is understood that $ \hat{U}_{x}=1$ if $x$ is a minimal element in $\calC$.
Given a polynomial of the free field at a point, $F(\hatphi_x)$, the Heisenberg operator, $F(\hatphi^H_x)$, is defined as,
\begin{equation}\label{eq: local heisenberg}
     F(\hatphi^H_x)=\hat{U}_{x}^{\dagger} F(\hatphi_x) \hat{U}_x.
\end{equation}
In particular we have $ \hatphi_x^H=\hat{U}_{x}^{\dagger} \hatphi_x \hat{U}_x$.
The S-operator is similarly defined as,
\begin{equation}
     \hat{S}=C\bigg[\prod_{x\in\calC} e^{-\frac{ig}{\hbar}V_x(\hatphi_x)}\bigg].
\end{equation}
The interacting vacuua are defined as $|0\rangle_\IN=|0\rangle$ and $|0\rangle_\out=\hat{S}^{\dagger}|0\rangle$, and similarly for particle states, $|\lambda_1\ldots \lambda_k\rangle_\IN=|\lambda_1\ldots \lambda_k\rangle$ and $|\lambda_1\ldots \lambda_k\rangle_\out=\hat{S}^{\dagger}|\lambda_1\ldots \lambda_k\rangle$. For an operator $A$ we write $\langle A\rangle_{\text{in-in}}:={}_\IN\langle0|A|0\rangle_\IN=\langle0|A|0\rangle$ and $\langle A\rangle_{\text{in-out}}:={}_\out\langle0|A|0\rangle_\IN=\langle0|\hat SA|0\rangle$.

The following lemma collects the causal factorisation properties of the Heisenberg fields used throughout. Besides the covariant evolution operator \eqref{eq: U operator}, it uses the label-ordered products
\[
\hat U^{(x)}:=C\Bigl[\prod_{z<x}e^{-\frac{ig}{\hbar}V_z(\hatphi_z)}\Bigr],\qquad
\hat U^{(x,y)}:=C\Bigl[\prod_{y\leq z<x}e^{-\frac{ig}{\hbar}V_z(\hatphi_z)}\Bigr]\quad(x>y),
\]
where the products run over labels; these are the operators $U_x$ and $U_{x,y}$ of \cite{Albertini:2024srq}.
\begin{lemma}[Causal factorisation]\label{lemma: causal factorisation}
Let $V$ be real. Coefficientwise in $g$:
\begin{enumerate}
\item[(i)] $\hatphi^H_x=\hat U_x^\dagger\hatphi_x\hat U_x=\hat U^{(x)\dagger}\hatphi_x\hat U^{(x)}$ for every $x\in\calC$;
\item[(ii)] $C[F(\hatphi^H)]=\hat S^\dagger\,C[\hat S\,F(\hatphi)]$ for every polynomial $F$, where on the right-hand side $C$ orders the factors $e^{-\frac{ig}{\hbar}V_z(\hatphi_z)}$ of $\hat S$ together with the fields;
\item[(iii)] $[\hatphi^H_x,\hatphi^H_y]=0$ if $x$ and $y$ are spacelike separated. In particular, causally ordered products of Heisenberg fields are well defined.
\end{enumerate}
\end{lemma}
\begin{proof}
Parts (i) and (iii) are \cite[eqs.~(3.9)--(3.11)]{Albertini:2024srq} and \cite[eq.~(3.22)]{Albertini:2024srq}, respectively.

For (ii), let $F(\hatphi)=\hatphi_{x_1}^{m_1}\cdots\hatphi_{x_k}^{m_k}$ be a monomial with distinct points, labelled so that $x_1>\cdots>x_k$, and write $\mathcal O(x_i)=\hatphi_{x_i}^{m_i}$. By \cite[Lemma~A.3 and eqs.~(3.7)--(3.8)]{Albertini:2024srq},
\[
\mathcal O^H(x_1)\cdots\mathcal O^H(x_k)=\hat U^{(x_1)\dagger}\mathcal O(x_1)\,\hat U^{(x_1,x_2)}\mathcal O(x_2)\cdots\hat U^{(x_{k-1},x_k)}\mathcal O(x_k)\,\hat U^{(x_k)} .
\]
Multiplying by $\hat S=C\bigl[\prod_{z\geq x_1}e^{-\frac{ig}{\hbar}V_z(\hatphi_z)}\bigr]\hat U^{(x_1)}$ gives the label-ordered product of all factors. This is $C[\hat S\,\mathcal O(x_1)\cdots\mathcal O(x_k)]$, because $e^{-\frac{ig}{\hbar}V_{x_i}(\hatphi_{x_i})}$ commutes with $\mathcal O(x_i)$. By (iii), label order agrees with causal order on Heisenberg fields, which gives (ii) for monomials and, by linearity, for all $F$.
\end{proof}
\begin{remark}
For $k=2$, inside the vacuum expectation value, (ii) is \cite[eq.~(5.24)]{Albertini:2024srq}. In the functional language of Section~\ref{Observables}, $C[\hat S\,F(\hatphi)]=\QT(\mathscr SF)$ by Definition~\ref{definition QT}.
\end{remark}

With the above construction, given any polynomial $F$ in the free field operators one can thus compute a vacuum in-in expectation value, a vacuum in-out correlation function and matrix elements of the various particle states. The perturbative diagrammatic expansions of each has been derived in \cite{Albertini:2024srq}.

\subsection{Path integral formalism}\label{sec:pathint}
The observables defined in the previous section using the operator formalism can be computed by a path integral as follows \cite{sorkin2011scalar,Albertini:2024srq}.

\begin{definition}[History, functional]\label{def history}\

  \begin{itemize}
\item A real scalar field \emph{history} or \emph{configuration} is a map $\phi:\calC\rightarrow \mathbb{R}$. The set of these is denoted $\RR^\calC$. For $z\in\calC$, denote $\phi_z=\phi(z)$.

\item A \emph{functional} is a smooth map from $\RR^\calC$ to $\CC$. A functional, $F$, is \emph{polynomial} if $F(\phi)$ is a (finite) linear combination of products of $\phi_z$'s, $z\in\calC$. We will denote the space of polynomial functionals as $\CC[\calC]$, since this is the free commutative algebra formally generated by the points of $\calC$.
\item A functional is \emph{local} if it is of the form 
\[
F(\phi) = \sum_{z\in\calC} F_z(\phi_z) ,
\]
for some smooth functions $F_z:\RR\to\CC$ for each $z\in\calC$.
\end{itemize}
\end{definition}

\begin{definition}\label{def formal S matrix}
Given some real local functional $V\in\CC[\calC]$,   the \emph{formal S-matrix} $\mathscr{S}$ is defined as a formal power series in $g$,
\begin{equation}\begin{split}
    \mathscr{S}(\phi):=&e^{-\frac{ig}{\hbar} V(\phi)}.
\end{split}
    \end{equation}
\end{definition}
\begin{remark} Some remarks regarding notation and conventions are in order:
\begin{enumerate}[(i)]
	\item    One can choose the coupling constant $g$ to vary from point to point by absorbing it into $V$ as $V=\sum_{z\in\calC} g_zV_z(\phi_z)$.
\item Note that $\mathscr S$ is a formal power series in $g$, but its coefficients are polynomial functionals.
\item The sign in $\mathscr S$ is opposite to that in \cite{Hawkins:2016ddq}. All formulae from there can be translated by replacing $g$ with $-\lambda$.
\item We write $\mathscr S^{-1}=e^{\frac{ig}{\hbar}V}$ for the pointwise inverse of $\mathscr S$; since $V$ is real, $\mathscr S^{-1}=\overline{\mathscr S}$. It should not be confused with the star inverses $\mathscr S^{{}^{-1,\star}}$ introduced in Section~\ref{sec:alg}.
\end{enumerate}
\end{remark}
The path-integral form of the in-in correlators on causal sets uses the histories formulation of quantum theory \cite{Sorkin:1994dt}. There, the dynamics and the initial state are not encoded in a probability measure on histories but in a decoherence functional $D$, which assigns a complex number to each \emph{pair} of histories. Heuristically, $D(\phi^+,\phi^-)$ is the amplitude of $\phi^+$ times the complex conjugate of the amplitude of $\phi^-$, summed over the final states, on which the two histories must agree. Correlators are obtained by integrating functions of the two histories against the decoherence functional. The resulting double path integral is of closed-time-path (Schwinger--Keldysh) form. Here $D_0$ is the decoherence functional of the free theory in the vacuum state $|0\rangle$: it is determined by $\Delta$ and $H=\Re W$, and for $g=0$ formula \eqref{eq: df master equation1} expresses the free in-in correlators in $|0\rangle$ as integrals against $D_0$. Given $F,G\in\CC[\calC]$, one can compute the in-in expectation value, which is defined in the operator formalism of Section~\ref{sec:op}, as a double path integral over pairs of histories,
\begin{align}\label{eq: df master equation1}
&\left\langle \overline{C}\left[ G(\hatphi^H) \right] C\left[ F(\hatphi^H) \right]\right\rangle_{\text{in-in
      }}=\int_{\mathbb{R}^{2N}}d^N\phi^-d^N\phi^+ \, G(\phi^-) \mathscr{S}^{-1}(\phi^-)D_0(\phi^+,\phi^-) \mathscr{S}(\phi^+) F(\phi^+)  \,
\end{align}
where the $C$ and $\overline{C}$ order the Heisenberg fields (\textit{e.g.} $C\left[ {\hatphi_x}^H {\hatphi_y}^H\right]={\hatphi_x}^H {\hatphi_y}^H$ if $x\not\prec y$),  $\phi^+$ and $\phi^-$ are two independent histories and $D_0(\phi^+,\phi^-)$ is the free decoherence functional. The left-hand side of \eqref{eq: df master equation1} is built from operators,
the equality to the integral on the right-hand side has been proven in \cite{albertini:2021,Jubb:2023mlv,Albertini:2024srq}. We provide an alternative proof of \eqref{eq: df master equation1} for general $F$ and $G$, coefficientwise in $g$, in Corollary~\ref{cor: in-in path integral} below. Following the Gaussian Fourier representation in \cite[Sec.~4]{Sorkin:2011pn}, we spell out here how to define the decoherence functional as a tempered distribution. Set
\[
u=\begin{pmatrix}\phi^-\\\phi^+\end{pmatrix},\qquad
\Sigma_0=\begin{pmatrix}\overline{\Delta^F}&W\\W^{\mathsf T}&\Delta^F\end{pmatrix}.
\]
With Fourier convention for $p\in\RR^{2N}$, $\widetilde f(p)=\int e^{ip^{\mathsf T}u}f(u)\,d^{2N}u$, define
\[
\widetilde D_0(p)\equiv\chi_0(p):=
\exp\!\left[-\frac{\hbar}{2}p^{\mathsf T}\Sigma_0p\right].
\]
The matrix $\Sigma_0$ is complex symmetric and
\[
\Re\Sigma_0=\begin{pmatrix}H&H\\H&H\end{pmatrix}\geq0,
\qquad H=\Re W.
\]
Thus, since $H\geq0$, $|\chi_0(p)|\leq1$ for real $p$, so its inverse Fourier transform defines a unique $D_0\in\mathcal S'(\mathbb R^{2N})$. $D_0$ naturally contains delta functions, which constrain the difference of the two histories. To see this, split $p=(p^-,p^+)$, and pass to the variables $P=p^-+p^+$ and $q=\frac12(p^--p^+)$, i.e. $p^\pm=\frac12P\mp q$. With the mean $M=\frac12(\phi^++\phi^-)$ and $s=\phi^--\phi^+$ one has $p^{\mathsf T}u=P^{\mathsf T}M+q^{\mathsf T}s$, so $P$ and $q$ are the Fourier variables dual to the mean and the difference of the two histories. Using \eqref{eq: W and DeltaF}, one finds
\[
p^{\mathsf T}\Sigma_0\,p=P^{\mathsf T}HP-2i\,P^{\mathsf T}\Delta^Rq = P^{\mathsf T}HP-2i\,q^{\mathsf T}\Delta^AP.
\]
In terms of the new variables, the inverse Fourier transform gives
\begin{align*}
D_0(\phi^+,\phi^-)&=\frac{1}{(2\pi)^{2N}}\int_{\RR^{2N}}d^{2N}p\;e^{-ip^{\mathsf T}u}\,\chi_0(p)\\
&=\frac{1}{(2\pi)^{2N}}\int_{\RR^N}d^NP\;e^{-iP^{\mathsf T}M-\frac\hbar2P^{\mathsf T}HP}\int_{\RR^N}d^Nq\;e^{-iq^{\mathsf T}(\phi^--\phi^+-\hbar\Delta^AP)},
\end{align*}
and since $\int_{\RR^N}d^Nq\,e^{-iq^{\mathsf T}v}=(2\pi)^N\delta^N(v)$,
\begin{equation}\label{eq: D0 explicit}
D_0(\phi^+,\phi^-)=\frac{1}{(2\pi)^N}\int_{\RR^N}d^NP\;e^{-iP^{\mathsf T}m-\frac\hbar2P^{\mathsf T}HP}\;\delta^N\!\bigl(s-\hbar\Delta^AP\bigr)
\end{equation}
in the sense of distributions. Written out componentwise and using $\Delta^A_{xy}=\Delta^R_{yx}$ (which vanishes unless $y\succ x$), the delta distribution in \eqref{eq: D0 explicit} imposes
\[
\phi^-_x-\phi^+_x=\hbar\sum_{y\succ x}\Delta^R_{yx}P_y\qquad\text{for all }x\in\calC .
\]
Thus the two histories can differ at $x$ only through the points in the future of $x$. If $x$ is a maximal element, the sum is empty, and $\phi^-_x=\phi^+_x$. Hence the two histories agree at every maximal element. This is the causal-set counterpart of the condition in the closed-time-path formalism that the two histories agree at the final time, cf.\ \cite[Sec.~4]{Sorkin:2011pn}.

\begin{remark}
Formula \eqref{eq: D0 explicit} also shows that $|D_0|$ has a symmetry. Shifting both histories by the same $w\in\RR^N$ leaves $\phi^--\phi^+$ unchanged and replaces $M$ by $M+w$, which enters \eqref{eq: D0 explicit} only through $e^{-iP^{\mathsf T}M}$. Setting $w=\Delta^Rc$, $c\in\RR^N$, and using the delta distribution to identify $\phi^--\phi^+=\hbar\Delta^AP$ we obtain
 \[
 P^{\mathsf T}\Delta^Rc=c^{\mathsf T}\Delta^AP=\frac1\hbar c^{\mathsf T}(\phi^--\phi^+)\,,
 \] 
 so
\[
D_0(\phi^++\Delta^Rc,\ \phi^-+\Delta^Rc)=e^{-\frac i\hbar c^{\mathsf T}(\phi^--\phi^+)}\,D_0(\phi^+,\phi^-).
\]
Hence $|D_0|$ is invariant under such translations. As a consequence, the integral of a polynomial against $D_0$ converges only as an oscillatory integral of Fresnel type. To evaluate such integrals one can use standard regularisation methods for oscillatory integrals.

For a polynomial $P$ in the two histories we define,
\[
\int d^{2N}u\,D_0(u)P(u):=\lim_{R\to\infty}\langle D_0,P\rho_R\rangle,
\]
where $\rho\in\mathcal S(\RR^{2N})$, $\rho(0)=1$ and $\rho_R(u)=\rho(u/R)$. Since every derivative of $\chi_0$ has at most polynomial growth, Fourier transformation and dominated convergence show that the limit exists, is independent of $\rho$, and
\begin{equation}\label{eq: moment formula}
\int d^{2N}u\,D_0(u)P(u)=\bigl[P(-i\partial_p)\chi_0(p)\bigr]_{p=0}.
\end{equation}
\end{remark}

The resulting distribution has the operator representation \cite{sorkin2011scalar}

 \begin{align}
    D_0(\phi^+,\phi^-) & = \left\langle \delta(\hatphi_1-\phi^-_1)\delta(\hatphi_2-\phi^-_2)\cdots \delta(\hatphi_N-\phi^-_N)\delta(\hatphi_N-\phi_N^+) \cdots\delta(\hatphi_2-\phi^+_2)\delta(\hatphi_1-\phi_1^+)\right\rangle \,.\label{eq:dcf0}
\end{align}
Upon integration, $D_0(\phi^+,\phi^-)$ causally orders functions of $\phi^+$ and anti-causally orders functions of $\phi^-$. 
Causally ordered in-out correlation functions are given by,
\begin{align}\label{eq: df master equation2}
 &     \left\langle C\left[ F(\hat{\phi}^H) \right]\right\rangle_{\text{in-out
      }}=\int_{\mathbb{R}^{2N}}d^N\phi^-d^N\phi^+ \, D_0(\phi^+,\phi^-) \mathscr{S}(\phi^+)F(\phi^+)  .
\end{align}
One obtains the following expression for particle matrix elements via the reduction formula \cite{Zalel:2026srq}, retaining all direct incoming--outgoing pairings and the particle normalisation used above (in \cite{Zalel:2026srq} the matrix elements are normalised by ${}_\out\langle0|0\rangle_\IN$; \eqref{eq: reduction} is the unnormalised version, proved in Proposition~\ref{prop: matrix element}):
\begin{equation}\label{eq: reduction}
\begin{split}
&{}_{\out}\left\langle\lambda_1\ldots\lambda_k\left|C[F(\hat\phi^H)]\right|\mu_1\ldots\mu_l\right\rangle_{\IN}\\
&\quad=\sum_{M\in\mathfrak M_{k,l}}c_M\hbar^{d_M/2}(-1)^{d_M}
\int d^N\phi^-d^N\phi^+\,
\bigl(\mathcal D_M^+D_0\bigr)(\phi^+,\phi^-)
\mathscr S(\phi^+)F(\phi^+).
\end{split}
\end{equation}
Here $\mathfrak M_{k,l}$ is the set of partial matchings between the outgoing slots $\{1,\ldots,k\}$ and incoming slots $\{1,\ldots,l\}$; each slot belongs to at most one pair. Let $O_M$ and $I_M$ be the unmatched outgoing and incoming slots, respectively, and set
\[
d_M=k+l-2|M|,\qquad
c_M=\prod_{(j,i)\in M}\frac{\delta_{\lambda_j\mu_i}}{\sqrt{\lambda_j\mu_i}},
\]
\begin{equation}\label{eq: def Dcal}
\mathcal D_M=
\prod_{j\in O_M}\left(\sum_{x\in\calC}\overline{w_x^{\lambda_j}}\,\partial_x\right)
\prod_{i\in I_M}\left(\sum_{x\in\calC}w_x^{\mu_i}\,\partial_x\right).
\end{equation}
The superscript $+$ means that these derivatives act on $\phi^+$ only. Slot labels distinguish repeated occurrences of the same mode. The empty matching recovers the fully attached term; nonempty matchings supply the direct particle overlaps.

\subsection{The algebraic formalism}\label{sec:alg}

AQFT on a causal set was formulated in \cite{Dable-Heath:2019sej}. Deformation quantisation has the advantage that the construction of the algebra of observables can be done completely abstractly, without the need for a distinguished state (\textit{e.g.}, a vacuum) and without invoking Fock space methods. At a later stage, one can then seek suitable states on the abstract algebra and use these to form Hilbert space representations by the GNS construction.

We begin with the free theory.

\begin{definition}
The \emph{support} of a functional $F:\RR^\calC\to\CC$ is
\[
\supp F = \{ x\in\calC \mid \exists \phi\in\RR^\calC,\ \tfrac{\partial F}{\partial \phi_x}\big|_{\phi}\neq0\} .
\]
\end{definition}
For a functional $F(\phi)$, we denote its derivatives as,
\[
\frac{\partial F}{\partial \phi_x}=\partial_x F = F_{,x}  .
\]

\begin{definition}
The  \emph{Moyal–Weyl product} $\star$ of $F,G\in\CC[\calC]$ is,
\begin{equation}
F\star G=\sum_{n=0}^\infty \frac{(i\hbar/2)^n}{n!} F_{,i_1\ldots i_n} \Delta_{i_1j_1}\ldots \Delta_{i_nj_n}G_{,j_1\ldots j_n},
\end{equation}
where $\Delta=2\Im W$ and we use the summation convention for the indices, which are valued in $\calC$.
\end{definition}

\begin{definition}
The  \emph{Wick product} $\starH$ of $F,G\in\CC[\calC]$ is,
\begin{equation}
F\starH G=\sum_{n=0}^\infty \frac{\hbar ^n}{n!} F_{,i_1\ldots i_n} W_{i_1j_1}\ldots W_{i_nj_n}G_{,j_1\ldots j_n}.
\end{equation}
\end{definition}
Note that both star products are associative. Because $F$ and $G$ are finite polynomials, the series are finite sums. This allows $\hbar>0$ to be fixed rather than a formal parameter. The products are non-commutative (since $\Delta$ and $W$ are not symmetric), although $F$ and $G$ commute if their supports are spacelike-separated (since $\Delta_{xy}=\Delta_{yx}=0$ and $W_{xy}=W_{yx}$ when $x$ and $y$ are spacelike separated).

The Moyal-Weyl product and the Wick product are related by the following ``gauge'' transformation.

\begin{definition} For $F\in \CC[\calC]$, let
\begin{equation}   
\alpha_H F:=e^{\frac{\hbar}{2}H^{ij}\partial^2_{ij}}F,
\end{equation}
where $H=\Re W$.
\end{definition}
This gauge transformation relates the two products via,
\begin{equation}
    \alpha_H F \starH \alpha_H G = \alpha_H (F\star G).
\end{equation}

In Section~\ref{Observables}, classical observables (functionals) are mapped to operators by quantisation maps. In particular, $\QH$ will be the quantisation map corresponding to normal (Wick) ordering with respect to the quasifree state determined by $W$. The gauge transformation $\alpha_H$ relates symmetric operator ordering to normal ordering.  Note that the former is state-independent and latter is not.

A state is a normalised positive linear functional on the algebra of observables.
\begin{definition}[The state $\omega_{H,0}$.]

\begin{equation}
    \omega_{H,0}(F)=F(0).
\end{equation}

\end{definition}

In this definition $F\in\mathfrak A_H(\calC)=(\CC[\calC],\starH)$. In the Moyal presentation the same state is $F\mapsto(\alpha_HF)(0)$. We illustrate the Wick presentation and its equivalence with the operator approach of Section \ref{sec:op} in the following example.

\begin{example}    The 2-point function is given by

\begin{equation}
        \begin{split}
            \omega_{H,0}(\phi_x\starH\phi_y)=(\phi_x\starH\phi_y)(0)=\hbar W_{xy},
        \end{split}
    \end{equation}
where the last equality follows from the definition of $\starH$.\end{example} 
  
Next we consider interactions.

\begin{definition}
The  \emph{time-ordered product} of  $F,G\in\CC[\calC]$ is, 
\begin{equation}\label{def: time ordered product}
F\dT G=\sum_{n=0}^\infty \frac{\hbar ^n}{n!} F_{,i_1\ldots i_n} \Delta^F_{i_1j_1}\ldots \Delta^F_{i_nj_n}G_{,j_1\ldots j_n}.
\end{equation}
\end{definition}

This product is associative and commutative. If  $x\not\prec y$ for all $x\in\supp F$ and $y\in\supp G$ then,
\begin{align}
F \starH G = F \dT G .
\end{align}
In this sense, $\dT$ is the time-ordered version of $\starH$.

\begin{definition}\label{def: starT}
The star product $\starT$ of $F,G\in\CC[\calC]$ is,
\begin{equation}
F\starT G=\sum_{n=0}^\infty \frac{\hbar ^n}{n!} F_{,i_1\ldots i_n} (-i\Delta^A_{i_1j_1})\ldots (-i\Delta^A_{i_nj_n})G_{,j_1\ldots j_n},
\end{equation}
where $\Delta^A$ is the advanced propagator.
\end{definition}
Note that unlike $\starH$ and $\dT$, the product $\starT$ depends only on the dynamics, not on the state: it depends on $W$ only through $\Delta=2\Im W$, which is common to all admissible two-point functions of the theory. Additionally, 
if $x\not\prec y$ for all $x\in\supp F$ and $y\in\supp G$, then only the first term in the expansion is non-vanishing and we obtain,
\begin{equation}
F \starT G = FG .
\end{equation}

In $\CC[\calC][[\hbar]]$, any $F=1+O(\hbar)$ has a unique inverse with respect to either star product, denoted by
\begin{equation}
    \begin{split}
        &F^{{}^{-1,\starH}}\starH F=1\\
        &F^{{}^{-1,\starT}}\starT F=1.
    \end{split}
\end{equation}

\begin{definition}
 The \emph{time-ordering operator} $\mathcal{T}:\CC[\calC]\to\CC[\calC]$  is,
\begin{equation}\begin{split}
  \mathcal{T}=&e^{\frac{\hbar}{2}\Delta^F_{ij}\partial^2_{ij}},
  \end{split}
\end{equation}
\textit{i.e.}, for $G\in\CC[\calC]$,
\begin{equation}\begin{split}
  \mathcal{T}G=\sum_{n=0}^\infty \frac{\hbar^n}{n! 2^n}\Delta^F_{i_1j_1}\ldots \Delta^F_{i_nj_n}G_{,i_1\ldots i_nj_1\ldots j_n}.
  \end{split}
\end{equation}
\end{definition}
\begin{remark}
Note that $\mathcal{T}$ and $\dT$ depend on the choice of Feynman propagator --- or equivalently the state or $H$. This was written as $\Tcal_H$
 in \cite{Hawkins:2016ddq}, but we won't write this dependence explicitly here.
\end{remark}

The inverse $\mathcal{T}^{-1}$ satisfies,
\begin{equation}\begin{split}
  \mathcal{T}( \mathcal{T}^{-1}G)=\mathcal{T}^{-1}( \mathcal{T}G)=G.
  \end{split}
\end{equation}
and is given explicitly by, \begin{equation}\begin{split}
  \mathcal{T}^{-1}=&e^{-\frac{\hbar}{2}\Delta^F_{ij}\partial^2_{ij}}.
  \end{split}
\end{equation}

The effect of $\mathcal{T}$ on a functional is to replace all pointwise products with time-ordered products. This  is expressed in the relation, \begin{equation}\label{time products relation}
\Tcal F\dT \Tcal G=\mathcal{T}(FG).
\end{equation}
Similarly, $\mathcal{T}$ intertwines between $\starH$ and $\starT$,
\begin{equation}\label{time ordering relation}
\Tcal F \starH \Tcal G = \Tcal(F\starT G) .
\end{equation}
We write $\overline{\Tcal}$ to denote the complex conjugate of $\Tcal$, namely $\overline{\Tcal}=e^{\frac{\hbar}{2}\overline{\Delta^F_{ij}}\partial^2_{ij}}.$
Below we give an example to help the reader gain some intuition.  Exponential identities in the following example are understood coefficientwise in an auxiliary formal parameter.

\begin{example}
First consider $F=\phi_x$. Since $\phi_x$ is linear we have $\mathcal{T}(\phi_x)=\phi_x$ by definition of $\mathcal{T}$. Now consider $F=\phi_x$ and $G=\phi_y$. By applying \eqref{time products relation} we obtain $\mathcal{T}(\phi_x\phi_y)= \phi_x\dT\phi_y.$
If $\mathcal{T}({\phi_{x_1} \cdots \phi_{x_k}})={\phi_{x_1}\dT \cdots \dT \phi_{x_k}}$, then by application of \eqref{time products relation} with $F={\phi_{x_1} \cdots \phi_{x_k}}$ and $G={\phi_{x_{k+1}}}$ we obtain $\mathcal{T}(\phi_{x_1} \cdots \phi_{x_{k+1}})= \phi_{x_1} \dT \cdots \dT \phi_{x_{k+1}}$, so by induction this holds for all $k$. When acting on an exponential of a polynomial $F$ we obtain, $\mathcal{T}e^F=\sum_{k=0}^{\infty}\frac{\mathcal{T}F^k}{k!}$, hence again all pointwise products are replaced with time-ordered products. Note also that $\mathcal{T}e^{\mathcal{T}^{-1}F}=\sum_{k=0}^{\infty}\frac{\mathcal{T}(\mathcal{T}^{-1}F)^k}{k!}=1+F+\frac{1}{2}F\dT F+\frac{1}{3!}F\dT F \dT F+\cdots$, so products between the $F$'s are replaced with time-ordered products, but products inside each $F$ remain pointwise.
\end{example}

For a classical observable $F$ the corresponding
quantum interacting field is given by $R_{H,gV}(F)$, where $R_{H,gV}$ is the retarded quantum M\o ller
operator defined  below. Physically, we interpret $R_{H,gV}(F)$, as the interacting observable corresponding to $F$, constructed from free ﬁelds.

\begin{definition}\label{def moller}
   For $V\in\CC[\calC]$, the M\o ller operators $R_{\mathcal{T},gV}$ and $R_{H, gV}$ are defined by, for $F\in\CC[\calC]$,
    \begin{equation}\begin{split}
&R_{\mathcal{T},gV}(F):=\mathscr{S}^{{}^{-1,\starT}}\starT \big(\mathscr{S} F\big),\\
&
R_{H, gV}(F):=\big(\mathcal{T}\mathscr{S}\big)^{{}^{-1,\starH}} \starH \bigg(\mathcal{T}\mathscr{S}\ \dT F\bigg),
\end{split}
\end{equation}
where $\mathscr{S}$ is the formal S-matrix as given in Definition \ref{def formal S matrix}.
\end{definition}
\begin{remark}
    Note that we define the M\o ller operators $R_{\mathcal{T}, gV}(F)$ and $R_{H, gV}(F)$ in the same way they were defined in \cite{Hawkins:2016ddq}. This is different to how they were defined in \cite{Dable-Heath:2019sej}.
\end{remark}
It follows from \eqref{time ordering relation} that these M\o ller operators are related via the time-ordering operator as,
 \begin{equation}\label{relation moller}
 R_{H,gV}(\mathcal{T} F)=\mathcal{T} R_{\mathcal{T},gV}(F).
 \end{equation}

\section{Observables from M\o ller operators}
\label{Observables}
In this section, we introduce (Def.~\ref{definition QH}) the quantisation map $\QH$, which is the main tool for connecting the algebraic formalism to the operator formalism.
We begin with several propositions that will be proven later in this section.
Our first proposition proves the equivalence of the free theory in the different approaches.

\begin{proposition}\label{proposition: free}
    For any $k\geq0$ and $x_1,\ldots,x_k \in\mathcal{C}$,

    \begin{equation}
            \langle 0|\hatphi_{x_1}\cdots  \hatphi_{x_k}|0\rangle = \phi_{x_1}\starH \cdots  \starH \phi_{x_k}(0).
    \end{equation}
\end{proposition}
\begin{remark}
    Note that the $x_i$ may or may not be distinct.
\end{remark}
The next proposition considers the special case where the free fields in the correlator appear in a particular order.
\begin{proposition}\label{proposition: T Tbar} For any $k,l\geq 0$ and $x_1,\ldots, x_k, y_1, \ldots, y_l\in \calC$,
\begin{align}   \langle 0|\overline{C}[\hatphi_{y_1}\cdots  \hatphi_{y_l}]C[\hatphi_{x_1}\cdots  \hatphi_{x_k}]|0\rangle=\overline{\mathcal{T}}( \phi_{y_1}\cdots\phi_{y_l})\starH\mathcal{T}( \phi_{x_1}\cdots\phi_{x_k})(0).
   \end{align}
   \end{proposition}
   Our next two propositions generalise Propositions \ref{proposition: free} and \ref{proposition: T Tbar} to the interacting theory.

   \begin{proposition}\label{proposition: in in unordered}
    For any $k\geq 0$ and $x_1,\ldots,x_k \in\mathcal{C}$,

    \begin{equation}
            _{\IN}  \langle 0|\hatphi^H_{x_1}\cdots  \hatphi^H_{x_k}|0\rangle _{\IN}  = R_{H, gV}(\phi_{x_1})\starH \cdots  \starH R_{H, gV}(\phi_{x_k})(0).
    \end{equation}
\end{proposition}

\begin{proposition}\label{proposition: T Tbar interacting} For any $k,l\geq 0$ and $x_1,\ldots, x_k, y_1, \ldots, y_l\in \calC$,
\begin{align}    _{\IN} \langle 0|\overline{C}[\hatphi^H_{y_1}\cdots  \hatphi^H_{y_l}]C[\hatphi^H_{x_1}\cdots  \hatphi^H_{x_k}]|0\rangle _{\IN} =\overline{\mathcal{T}}(\mathscr{S}^{-1} \phi_{y_1}\cdots\phi_{y_l})\starH\mathcal{T}( \mathscr{S}\phi_{x_1}\cdots\phi_{x_k})(0).
   \end{align}
   \end{proposition}

   The next two propositions give the causally-ordered in-in and in-out correlators, respectively.
   \begin{proposition}\label{proposition: in in ordered} For any $k\geq 0$ and $x_1,\ldots,x_k \in\mathcal{C}$,

\begin{align}
      \label{proposition: in in}    _{\IN} \langle 0|C[{\hatphi_{x_1}^H}\cdots\hatphi_{x_k}^H]|0\rangle_{\IN}&=R_{H, gV}(\mathcal{T}(\phi_{x_1}\cdots\phi_{x_k}))(0).
   \end{align}
   \end{proposition}
   
\begin{proposition}\label{proposition: equivlence} For any $k\geq 0$ and $x_1,\ldots,x_k \in\mathcal{C}$,

\begin{align}
      \label{proposition: in out}    _{\out} \langle 0|C[{\hatphi_{x_1}^H}\cdots\hatphi_{x_k}^H]|0\rangle_{\IN}&=\mathcal{T}( \mathscr{S}\phi_{x_1}\cdots\phi_{x_k})(0).
   \end{align}

   \end{proposition}
Often one uses the normalized version of the in-out correlator, \textit{i.e.}

\begin{align} \label{proposition: in out normalised}
_{\out} \langle C[{\hatphi_{x_1}^H}\cdots\hatphi_{x_k}^H] \rangle^{\textrm{norm}}_{\IN}=
  \frac{_{\out} \langle 0|C[{\hatphi_{x_1}^H}\cdots\hatphi_{x_k}^H]|0\rangle_{\IN}}{ _{\out} \langle 0|0\rangle_{\IN}}&=\frac{\mathcal{T}( \mathscr{S}\phi_{x_1}\cdots\phi_{x_k})(0)}{\mathcal{T}( \mathscr{S})(0)}.
   \end{align}

\begin{proposition}\label{prop: matrix element} For $k,l,n\geq0$ and $x_1,\ldots,x_n\in\calC$,
\begin{equation}
\begin{split}
&{}_{\out}\langle\lambda_1\ldots\lambda_k|
C[\hat\phi^H_{x_1}\cdots\hat\phi^H_{x_n}]
|\mu_1\ldots\mu_l\rangle_{\IN}\\
&\quad=\sum_{M\in\mathfrak M_{k,l}}c_M\hbar^{d_M/2}
\left[\Tcal\mathcal D_M
\left(\mathscr S\phi_{x_1}\cdots\phi_{x_n}\right)\right](0),
\end{split}
\end{equation}
with the partial matchings $M$, the constants $c_M$, $d_M$ and the differential operators $\mathcal D_M$ defined in \eqref{eq: def Dcal}.
\end{proposition}
For $k=l=0$ only the empty matching contributes, with $c_\emptyset=1$ and $d_\emptyset=0$, and Proposition~\ref{prop: matrix element} reduces to Proposition~\ref{proposition: equivlence}. The path-integral form \eqref{eq: reduction} of this result is Corollary~\ref{cor: reduction} below.

\subsection{Quantisation maps}
To prove Propositions \ref{proposition: free}--\ref{proposition: equivlence}, we introduce quantisation maps in order to compare the algebraic and operator formalisms. Each quantisation map corresponds to a choice of operator ordering.
Recall that $\CC[\calC]$ denotes the space of polynomial functionals. Let $\Op$ denote the space of polynomials in the free field operators.

\begin{definition}\label{definition QT}
The \emph{time-ordered quantisation map} $\QT:\CC[\calC]\rightarrow\Op$ is defined by $\QT(\phi_z)=\hatphi_z$ for all $z\in\calC$ and replacing pointwise polynomial products with causally ordered products. The \emph{anti time-ordered quantisation map} $\QTbar:\CC[\calC]\rightarrow\Op$ is similarly defined by $\QTbar(\phi_z)=\hatphi_z$ for all $z\in\calC$ and replacing pointwise polynomial products with anti-causally ordered products.
\end{definition}
\begin{remark}
In general, a causally ordered product of operators is not defined, because operators don't have a notion of support on $\calC$, but $\dT$ is the corresponding construction in the functional formalism, which is well-defined because this is off-shell.  For products of field operators, the causal orderings of Section~\ref{sec:op} are given by the quantisation maps: $C[\hatphi_{x_1}\cdots\hatphi_{x_k}]=\QT(\phi_{x_1}\cdots\phi_{x_k})$ and $\overline C[\hatphi_{x_1}\cdots\hatphi_{x_k}]=\QTbar(\phi_{x_1}\cdots\phi_{x_k})$.
\end{remark}

\begin{example} 
Let $x\not\prec y$, $x\not\prec z$ and $y\not\prec z$. Then,

\begin{align*}&\QT(\phi_y^3\phi_x\phi_z^2)= \hat{\phi}_x\hat{\phi}_y^3\hat{\phi}_z^2, \ \ \QT(\mathscr{S})= \hat{S}, \ \ \QT(\mathscr{S}\phi_y^3\phi_x\phi_z^2)= C[\hat{S}\hatphi_y^3\hatphi_x\hatphi_z^2],  \\ &\QTbar(\phi_y^3\phi_x\phi_z^2)=\hat{\phi}_z^2\hat{\phi}_y^3\hat{\phi}_x,  \ \ \QTbar(\overline{\mathscr{S}})= \hat{S}^{\dagger}, \  \QTbar(\overline{\mathscr{S}}\phi_y^3\phi_x\phi_z^2)= \overline{C}[\hat{S}^{\dagger}\hatphi_y^3\hatphi_x\hatphi_z^2].
\end{align*}

\end{example}

\begin{lemma}
For all $F,G\in\CC[\calC]$,
\begin{equation}\label{QT intertwines}
\QT(F)\QT(G) = \QT(F\starT G) .
\end{equation}
\end{lemma}
\begin{proof}
If $F$ is a constant, then this is trivial, since $\QT(1)=1$ and $1\starT G=G$.

Keeping in mind that $\QT(G)$ is, by definition, a linear combination of causally ordered products of field operators, consider $\QT(\phi_x)\QT(G)$ for some $x\in\calC$. To relate this to a causally ordered product, we need to move $\QT(\phi_x)=\hatphi_x$ past any $\hatphi_y$ with $x\prec y$. Since $[\hatphi_x,\hatphi_y]=i\hbar\Delta_{xy}$, this gives
\begin{align*}
\QT(\phi_x)\QT(G) &= \QT(\phi_xG) + \sum_{y\succ x}i\hbar\Delta_{xy}\QT(G_{,y})
= \QT(\phi_xG) - \sum_{y\in\calC}i\hbar\Delta^A_{xy}\QT(G_{,y})
= \QT(\phi_x\starT G) .
\end{align*}
This proves eq.~\eqref{QT intertwines} for $\deg F=1$.

Now suppose that eq.~\eqref{QT intertwines} is true for $\deg F\leq n$. Let $F$ be a monomial of degree $n+1$. This can be written as $F(\phi)=F'(\phi)\phi_x$, where $\phi_x$ is the first factor by label, so that $F'\starT\phi_x = F'\phi_x$. This gives
\begin{align*}
\QT(F)\QT(G) &= \QT(F'\starT \phi_x)\QT(G) = \QT(F')\hatphi_x\QT(G) = \QT(F') \QT(\phi_x\starT G) \\
&= \QT(F'\starT \phi_x \starT G) = \QT(F\starT G) ,
\end{align*}
using the assumption and associativity. By linearity, this implies eq.~\eqref{QT intertwines} for all polynomial $F$ of degree $\leq n+1$. By induction, this proves eq.~\eqref{QT intertwines} for all $F$.
\end{proof}
\begin{remark}
Equation~\eqref{QT intertwines} would usually be an asymptotic relation, involving a family of maps parametrized by $\hbar$, but with polynomial functionals it is an exact relation. 
\end{remark}

\begin{definition}\label{definition QH}
$\QH:=\QT\circ\Tcal^{-1}$
\end{definition}
\begin{lemma}\label{lemma:Q intertwines}
    \begin{equation} \label{Q intertwines}
\QH(F)\QH(G) = \QH(F\starH G) .
\end{equation}
\end{lemma}
\begin{proof}
    In \eqref{QT intertwines}, replace the $\QT$'s with $\QH$'s using definition \ref{definition QH}. Then use \eqref{time ordering relation} to rewrite the RHS and the result follows.
\end{proof}
\begin{lemma}\label{Expectation lemma}
For any $F\in\CC[\calC]$ and $x_1,\ldots,x_m\in\calC$,
\begin{align}
\QH(\overline F)&=\QH(F)^\dagger,\label{eq QH adjoint}\\
\QH(\phi_{x_1}\cdots\phi_{x_m})&={:}\hatphi_{x_1}\cdots\hatphi_{x_m}{:}\,,\label{eq normal ordering}
\end{align}
that is, $\QH$ is normal ordering with respect to the Fock vacuum, and
\begin{equation}\label{eq expectation}
\langle0\rvert\QH(F)\lvert0\rangle = F(0)
\end{equation}
\end{lemma}
\begin{proof}
Firstly, since $\Tcal^{-1}\phi_x=\phi_x$,
\[
\QH(\phi_x) = \hat\phi_x .
\]
Next, let $J$ be a formal $n$-dimensional vector. By eq.~\eqref{Q intertwines},
\[
e^{J\cdot \hat\phi} = e^{\QH(J\cdot\phi)} = \QH(\exp_{\starH}J\cdot\phi) \ .
\]
The $\starH$-exponential of this linear functional is
\[
\exp_{\starH}J\cdot\phi = e^{\frac\hbar2 J\cdot HJ} e^{J\cdot\phi} ,
\]
so
\begin{equation}\label{eq:normal-generating}
\QH(e^{J\cdot\phi})=e^{-\frac\hbar2 J\cdot HJ}e^{J\cdot\hat\phi}.
\end{equation}
Since $H$ is real, treating $J$ as real, taking adjoints, and comparing coefficients in $J$ proves eq.~\eqref{eq QH adjoint}.

Next, still treating $J$ as real, $J\cdot \hat\phi = A + A^\dag$, where
\[
A = \sum_{\lambda} \sqrt{\hbar\lambda}(J\cdot w^\lambda)a_\lambda \ .
\]
Using the commutator,
\[
[A,A^\dag] = \sum_{\lambda,\lambda'} \hbar\sqrt{\lambda\lambda'}(J\cdot w^\lambda)(J\cdot \overline w^{\lambda'})[a_\lambda,a^\dag_{\lambda'}] = \sum_\lambda \hbar\lambda (J\cdot w^\lambda)(J\cdot\overline w^\lambda) = \hbar J\cdot WJ = \hbar J\cdot H J ,
\]
the Campbell--Baker--Hausdorff identity gives $e^{J\cdot\hat\phi} = e^{A^\dag}e^A e^{\frac\hbar2 J\cdot HJ}$, therefore by eq.~\eqref{eq:normal-generating},
\begin{equation}\label{normal order}
\QH(e^{J\cdot\phi}) = e^{A^\dag}e^A =:e^{J\cdot\hat\phi}:,
\end{equation}
and comparing coefficients of $J$ proves eq.~\eqref{eq normal ordering} and
\[
\langle0\rvert \QH(e^{J\cdot\phi}) \lvert0\rangle = 1
\]
Again, comparing coefficients, we obtain
\[
\langle0\rvert \QH (\phi_{x_1}\dots\phi_{x_k}) \lvert0\rangle = 0
\]
for $k\geq1$. Equation~\eqref{eq expectation} follows by linearity.

\end{proof}

\begin{lemma}\label{lemma QTbar}
\begin{equation}
      \QH=\QTbar \circ\overline{\Tcal}^{\ -1}.
\end{equation}
  \end{lemma}
\begin{proof}
The adjoint of a causally ordered product is anti-causally ordered, so $\QTbar(F)=\QT(\overline F)^\dagger$. Rearranging the definition of $\QH$ and using eq.~\eqref{eq QH adjoint} now gives
\[
\QTbar(F)=\QT(\overline F)^\dagger
=\QH(\Tcal\overline F)^\dagger
=\QH(\overline\Tcal F) . \qedhere
\]
\end{proof}

\subsection{Proofs of the propositions}
\begin{proof}[Proof of Proposition \ref{proposition: free}] Using definition \ref{definition QH} of $\QH$ and the relation \eqref{Q intertwines} we obtain, 
 
\begin{equation}
            \langle 0|\hatphi_{x_1}\cdots  \hatphi_{x_k}|0\rangle =  \langle 0|\QH(\phi_{x_1})\cdots  \QH(\phi_{x_k})|0\rangle= \langle 0|\QH(\phi_{x_1}\starH \cdots  \starH \phi_{x_k})|0\rangle=\phi_{x_1}\starH \cdots  \starH \phi_{x_k}(0),
    \end{equation}

    where the last equality follows from Lemma \ref{Expectation lemma}.
\end{proof}

\begin{proof}[Proof of Proposition \ref{proposition: T Tbar}] By  definitions \ref{definition QT} and \ref{definition QH} of the quantisation maps we obtain,
\begin{equation}
    \begin{split}
        \langle 0|\overline{C}[\hatphi_{y_1}\cdots  \hatphi_{y_l}]C[\hatphi_{x_1}\cdots  \hatphi_{x_k}]|0\rangle&=\langle 0|\QTbar(\phi_{y_1}\cdots  \phi_{y_l})\QT(\phi_{x_1}\cdots  \phi_{x_k})|0\rangle\\
&=\langle 0|\QH \overline{\Tcal}(\phi_{y_1}\cdots  \phi_{y_l})\QH \Tcal(\phi_{x_1}\cdots  \phi_{x_k})|0\rangle,
    \end{split}
\end{equation}  
where to get the last line we applied Lemma \ref{lemma QTbar}. 
Applying Lemma \ref{lemma:Q intertwines} and then Lemma \ref{Expectation lemma} completes the proof.
\end{proof}

\begin{proof}[Proof of Proposition \ref{proposition: in in unordered}]
Lemma~\ref{lemma: causal factorisation}(ii) with $F=\phi_x$, together with, $\hat S^\dagger=\QTbar(\mathscr S^{-1})$ (as $V$ is real) and Lemma~\ref{lemma QTbar}, give
\begin{equation}\label{eq:0709}
\begin{split}
\hat\phi^H_{x_1}\cdots\hat\phi^H_{x_k}
&=\hat S^\dagger C[\hat S\hat\phi_{x_1}]\cdots
\hat S^\dagger C[\hat S\hat\phi_{x_k}]\\
&=\QH\!\left(
\overline\Tcal(\mathscr S^{-1})\starH\Tcal(\mathscr S\phi_{x_1})
\starH\cdots\starH
\overline\Tcal(\mathscr S^{-1})\starH\Tcal(\mathscr S\phi_{x_k})
\right).
\end{split}
\end{equation}
The required inverse identity is proved directly in the functional algebra. Put $A_z=\Tcal e^{-igV_z/\hbar}$. At a single site $\starH$ agrees with $\dT$, so $A_z^{-1,\starH}=\Tcal e^{igV_z/\hbar}$. Causal factorisation gives $\Tcal\mathscr S=A_N\starH\cdots\starH A_1$. Since $\Delta^F_{zz}$ is real, the anti-causal factorisation of $\overline\Tcal\mathscr S^{-1}$ is the reversed product of these inverses. Hence
\begin{equation}\label{eq: unitarity}
\overline\Tcal\mathscr S^{-1}\starH\Tcal\mathscr S=1,
\qquad
\overline\Tcal\mathscr S^{-1}=(\Tcal\mathscr S)^{-1,\starH}.
\end{equation}
This does not assume that $\QH$ is injective. Use \eqref{eq: unitarity} in \eqref{eq:0709}, apply Definition~\ref{def moller} and then Lemma~\ref{Expectation lemma} to obtain the result.
\end{proof}

\begin{proof}[Proof of Proposition \ref{proposition: T Tbar interacting}] By Lemma~\ref{lemma: causal factorisation}(ii), $C[F(\hatphi^H)]=\hat S^\dagger C[\hat SF(\hatphi)]=\hat S^\dagger\QT(\mathscr SF)$ and, taking adjoints, $\overline C[G(\hatphi^H)]=\bigl(C[\overline G(\hatphi^H)]\bigr)^\dagger=\QTbar(\mathscr S^{-1}G)\,\hat S$. Hence,
    \begin{equation}
        \begin{split}
            _{\IN}  \langle 0|   \overline{C}[\hatphi^H_{y_1}\cdots  \hatphi^H_{y_l}]C[\hatphi^H_{x_1}\cdots  \hatphi^H_{x_k}] |0\rangle _{\IN} &= \langle 0| \overline{C}[\hat{S}^{-1}\hatphi_{y_1}\cdots\hatphi_{y_l}]C[\hat{S}\hatphi_{x_1}\cdots  \hatphi_{x_k}]|0\rangle.
        \end{split}
    \end{equation} 
    The result follows from Proposition \ref{proposition: T Tbar}.
\end{proof}

\begin{proof}[Proof of Proposition \ref{proposition: in in ordered}]
By Proposition \ref{proposition: T Tbar interacting},

\begin{align}
      \label{eq:in-in-proof}    _{\IN} \langle 0|C[{\hatphi_{x_1}^H}\cdots\hatphi_{x_k}^H]|0\rangle_{\IN}&=\overline{\mathcal{T}}(\mathscr{S}^{-1})\starH\mathcal{T}( \mathscr{S}\phi_{x_1}\cdots\phi_{x_k})(0).
   \end{align}

Applying the unitarity argument of \eqref{eq: unitarity} and then definition \ref{def moller} of $R_{H,gV}$ gives the result.
   \end{proof}
   
\begin{proof}[Proof of Proposition \ref{proposition: equivlence}]
  
\begin{equation}\begin{split}
       _{\out} \langle 0|C[{\hatphi_{x_1}^H}\cdots\hatphi_{x_k}^H]|0\rangle_{\IN}=&   \langle 0| C[\hat{S}{\hatphi_{x_1}}\cdots\hatphi_{x_k}]|0\rangle      =\mathcal{T}(\mathscr{S}\phi_{x_1}\cdots\phi_{x_k})(0),
       \end{split}
   \end{equation}
   where the first equality follows from ${}_\out\langle0|=\langle0|\hat S$ and Lemma~\ref{lemma: causal factorisation}(ii), and the last equality follows from Proposition \ref{proposition: T Tbar}.
\end{proof}

\begin{proof}[Proof of Proposition \ref{prop: matrix element}]
We use the mode functions from \eqref{W decomposition} to define 
\[
\Phi^{\out}:=\prod_{j=1}^k\phi\cdot \overline{w^{\lambda_j}},\qquad
\Phi^{\IN}:=\prod_{i=1}^l\phi\cdot w^{\mu_i}.
\]
It follows from the definition of $\hat{\phi}(f)$ given in \eqref{Field definition} and the fact that $\QH$ imposes normal ordering that
\begin{equation}\label{eq: states from modes 1}
\QH(\Phi^{\IN})|0\rangle=\prod_i\sqrt{\hbar\mu_i}\,a^\dagger_{\mu_i}|0\rangle=\hbar^{l/2}\Bigl(\prod_i\mu_i\Bigr)|\mu_1\ldots\mu_l\rangle\,,
\end{equation}
and similarly
\begin{equation}\label{eq: states from modes 2}
\langle0|\QH(\Phi^{\out})=\hbar^{k/2}\Bigl(\prod_j\lambda_j\Bigr)\langle\lambda_1\ldots\lambda_k|
\end{equation}
Note that in \eqref{eq: states from modes 1}, the contributions involving annihilation operators vanish, since these are brought to the right and hit the vacuum vector $|0\rangle$. Similarly, the contributions from creation operators do not appear in \eqref{eq: states from modes 2}.

Since ${}_\out\langle\lambda_1\ldots\lambda_k|=\langle\lambda_1\ldots\lambda_k|\hat S$, Lemma~\ref{lemma: causal factorisation}(ii) implies
\[
{}_{\out}\langle\lambda_1\ldots\lambda_k|C[\hatphi^H_{x_1}\cdots\hatphi^H_{x_n}]|\mu_1\ldots\mu_l\rangle_{\IN}
=\langle\lambda_1\ldots\lambda_k|\,\QT(\mathscr S\phi_{x_1}\cdots\phi_{x_n})\,|\mu_1\ldots\mu_l\rangle .
\]
Using \eqref{eq: states from modes 1}, \eqref{eq: states from modes 2}, Lemma~\ref{lemma:Q intertwines} and Lemma~\ref{Expectation lemma}, we obtain
\begin{equation}\label{eq: matrix element star}
\langle\lambda_1\ldots\lambda_k|\,\QT( \mathscr S\phi_{x_1}\cdots\phi_{x_n})\,|\mu_1\ldots\mu_l\rangle
=\frac{\bigl(\Phi^{\out}\starH(\Tcal \mathscr S\phi_{x_1}\cdots\phi_{x_n})\starH\Phi^{\IN}\bigr)(0)}{\hbar^{(k+l)/2}\prod_j\lambda_j\prod_i\mu_i}.
\end{equation}
Now we compute star products with mode functionals. Since $W^{\mathsf T}\overline{w^\lambda}=\overline{Ww^\lambda}=\lambda\,\overline{w^\lambda}$ and $Ww^\mu=\mu\,w^\mu$, the definition of $\starH$ gives, for every functional $F$,
\[
\phi\bigl(\overline{w^\lambda}\bigr)\starH F=\phi\bigl(\overline{w^\lambda}\bigr)F+\hbar\lambda\,\mathcal D^{\out}_\lambda F,\qquad
F\starH\phi\bigl(w^\mu\bigr)=F\,\phi\bigl(w^\mu\bigr)+\hbar\mu\,\mathcal D^{\IN}_\mu F\,,
\]
where
\[
\mathcal D^{\out}_\lambda:=\sum_{x\in\calC}\overline{w^\lambda_x}\,\partial_x,\qquad
\mathcal D^{\IN}_\mu:=\sum_{x\in\calC}w^\mu_x\,\partial_x,
\qquad\text{so that}\qquad
\mathcal D_M=\prod_{j\in O_M}\mathcal D^{\out}_{\lambda_j}\prod_{i\in I_M}\mathcal D^{\IN}_{\mu_i}.
\]
Thus
\[
\Phi^{\out}\starH F=\sum_{O\subseteq\{1,\ldots,k\}}\ \prod_{j\notin O}\phi\bigl(\overline{w^{\lambda_j}}\bigr)\prod_{j\in O}\hbar\lambda_j\,\mathcal D^{\out}_{\lambda_j}F,
\qquad
F\starH\Phi^{\IN}=\sum_{I\subseteq\{1,\ldots,l\}}\Bigl(\prod_{i\in I}\hbar\mu_i\,\mathcal D^{\IN}_{\mu_i}F\Bigr)\prod_{i\notin I}\phi\bigl(w^{\mu_i}\bigr).
\]
and
\[
\bigl(\Phi^{\out}\starH(\Tcal \mathscr S\phi_{x_1}\cdots\phi_{x_n})\starH\Phi^{\IN}\bigr)(0)
=\sum_{M\in\mathfrak M_{k,l}}\ \prod_{j=1}^k\hbar\lambda_j\prod_{i\in I_M}\hbar\mu_i\prod_{(j,i)\in M}\delta_{\lambda_j\mu_i}\;\bigl[\mathcal D_M\Tcal \mathscr S\phi_{x_1}\cdots\phi_{x_n}\bigr](0).
\]
After combining all the constant, we obtain the desired result

\end{proof}

\subsection{Examples}\label{sec: examples}
We give some examples of how one can compute correlators in practice using the algebraic approach. It will be useful to pursue the analogy between the algebraic and the operator approach, as follows. By analogy with the evolution operator \eqref{eq: U operator} we define,

%First, it will be useful to recall the analogy between the algebraic and operator pictures: the M\o ller operator takes the place of the Heisenberg operator, the product $\starH$ takes the place of the operator product (cf.~relation \eqref{Q intertwines}). We can see this analogy when considering $R_{H,gV}(F_x)$ for a local functional $F_x=F_x(\phi_x)$. Define (cf.~\eqref{eq: U operator}),

\begin{definition}\label{definition U} For $x \in\calC$,
    \begin{align}
        %&\mathscr{U}_{x,y}:=\mathcal{T}(e^{-\frac{i}{\hbar}gV_{x-1}}\cdots e^{-\frac{i}{\hbar}gV_y})=\mathcal{T}e^{-\frac{i}{\hbar}gV_{x-1}}\starH\cdots \starH\mathcal{T}e^{-\frac{i}{\hbar}gV_y},\\
        &\label{def U}\mathscr{U}_{x}:=\mathcal{T}\biggl(\prod_{y\prec x}e^{-\frac{i}{\hbar}gV_{y}(\phi_y)}\biggr).
    \end{align}
\end{definition}
The following lemma shows that the M\o ller operator can be written in the same conjugation form as that of local Heisenberg operators, cf.~\eqref{eq: local heisenberg},
\begin{lemma}\label{lemma: conjugation} Given a polynomial $F$ with $\supp F\subseteq\{x\}$,
    \begin{align}
        &\label{lemma U 1}R_{H,gV}(\mathcal{T}F)=\mathscr{U}^{{}^{-1,\starH}}_x \starH \mathcal{T}F \starH \mathscr{U}_x.
        \end{align}
\end{lemma}
\begin{proof}
    Starting from definition \ref{def moller} of $R_{H,gV}$, and applying \eqref{time products relation} we obtain,
    \begin{equation}
        \begin{split}
            R_{H,gV}(\mathcal{T}F)=\big(\mathcal{T}\mathscr{S}\big)^{{}^{-1,\starH}} \starH \mathcal{T}(\mathscr{S}F).
        \end{split}
    \end{equation}
Now, writing $\mathscr{S}F=\bigg(\prod_{z\not\prec x}e^{-\frac{i}{\hbar}gV_{z}(\phi_z)}\bigg) F \bigg(\prod_{z\prec x} e^{-\frac{i}{\hbar}gV_{z}(\phi_z)}\bigg)$ and applying relation \eqref{time products relation} and definition \ref{definition U}, we can rewrite the RHS of the star product as,  \begin{equation}
    \mathcal{T}(\mathscr{S}F)=\Tcal\bigg(\prod_{z\not\prec x}e^{-\frac{i}{\hbar}gV_{z}(\phi_z)}\bigg)\starH \Tcal F \starH \mathscr{U}_{x}.
\end{equation}
Similarly, writing $\mathcal{T}\mathscr{S}=\Tcal\bigg(\prod_{z\not\prec x}e^{-\frac{i}{\hbar}gV_{z}(\phi_z)}\bigg)\starH \mathscr{U}_{x}$ and applying the definition of the star inverse, we obtain the relation,
\begin{equation}
\big(\mathcal{T}\mathscr{S}\big)^{{}^{-1,\starH}}\starH\Tcal\bigg(\prod_{z\not\prec x}e^{-\frac{i}{\hbar}gV_{z}(\phi_z)}\bigg)=\mathscr{U}^{{}^{-1,\starH}}_x.
\end{equation}This completes the proof.\end{proof}
Noting that $\mathscr{U}_x=1$ whenever $x$ is a minimal element, the following corollary is immediate.
\begin{corollary}\label{cor: minimal}
    $R_{H,gV}(\Tcal F)=\Tcal F$ when $x$ is a minimal element in $\calC$.
\end{corollary}

\begin{example} Consider the causal set $\calC=\{1,2\}$ with $1\prec 2$. We will compute the in-in expectation value $\langle \hatphi_2^H\hatphi_1^H \rangle$ in $\phi^4$ theory.
Starting from Proposition \ref{proposition: in in unordered}, and applying Lemma \ref{lemma: conjugation} and Corollary \ref{cor: minimal}, we have,
\begin{equation}
  \langle \hatphi_2^H\hatphi_1^H \rangle=\mathscr{U}^{{}^{-1,\starH}}_2\starH\phi_2\starH \mathscr{U}_2\starH  \phi_1(0).
\end{equation}
Multiply out the $\starH$ products to obtain,
\begin{equation}\label{eq: 12081}
\begin{aligned}
 \langle \hatphi_2^H\hatphi_1^H \rangle
  =\mathscr{U}^{{}^{-1,\starH}}_2\starH (\phi_2 \mathscr{U}_2+\hbar W_{21}\partial_1 \mathscr{U}_2)\starH  \phi_1(0)=\bigg(\phi_2+i\hbar\Delta^R_{21}(\mathscr{U}^{{}^{-1,\starH}}_2\starH \partial_1 \mathscr{U}_2 )\bigg)\starH  \phi_1(0).
   \end{aligned}
    \end{equation}
Now evaluate $\partial_1\mathscr{U}_2$ as,
\begin{equation}
\begin{aligned}
    \partial_1\mathscr{U}_2 &=-\frac{i}{\hbar}g\mathcal{T}(e^{-\frac{i}{\hbar}gV_1}\partial_1V_1)
    =-\frac{i}{\hbar}g\mathcal{T}e^{-\frac{i}{\hbar}gV_1}\starH\mathcal{T}\partial_1V_1=
        -\frac{ig}{\hbar}\mathscr{U}_2 \starH\mathcal{T}\partial_1V,
\end{aligned}
\end{equation}
and note that therefore,
\begin{equation}
\mathscr{U}^{{}^{-1,\starH}}_2\starH \partial_1 \mathscr{U}_2 = -\frac{ig}{\hbar}\mathcal{T}\partial_1V.
\end{equation}
Plug this back into \eqref{eq: 12081} and set $V=\sum_x\frac{1}{4!}\phi_x^4$ to obtain the result,
\begin{equation}\begin{split}
  \langle \hatphi_2^H\hatphi_1^H \rangle
  =\bigg(\phi_2+\frac{g}{3!}\Delta^R_{21}(\phi_1^3+3\hbar\Delta^F_{11}\phi_1)\bigg)\starH  \phi_1(0)=\hbar \Delta^F_{21}+\frac{\hbar^2g}{2}\Delta^R_{21}(\Delta^F_{11})^2.
   \end{split}
    \end{equation}
\end{example}
\begin{example} Consider the causal set $\calC=\{1,2,3\}$ with $1\prec 2\prec 3$. We will compute the in-in expectation value $\langle \hatphi_3^H \rangle$ in $\phi^3$ theory. 
Starting from Proposition \ref{proposition: in in unordered} and applying relation \eqref{relation moller}, we have,
\begin{equation}
    \langle \hatphi_3^H \rangle=\mathcal{T}R_{\mathcal{T},gV}(\phi_3)(0).
\end{equation}
First we compute $R_{\mathcal{T},gV}(\phi_3)$. Starting from definition \ref{def moller} of $R_{\mathcal{T},gV}$, expand the $\starT$ product as a series in $\hbar$ (using definition \ref{def: starT}). Grouping the terms by the number of derivatives acting on $\phi_3$, we obtain,
\begin{equation}\label{eq: derivation 2}
    R_{\mathcal{T},gV}(\phi_3)=\phi_3-i\hbar\Delta^R_{3j}\ \partial_j\mathscr{S}^{{}^{-1,\starT}}\starT\mathscr{S},\\
\end{equation}
where we use the shorthand $\partial_j\equiv\frac{\partial}{\partial\phi_j}$.
 To evaluate the second term on the RHS, note that expanding $\partial_j(\mathscr{S}^{{}^{-1,\starT}}\starT\mathscr{S})=0$ by the Leibniz rule gives,
\begin{equation}\begin{split}\label{eq: derivation 1}
\partial_j\mathscr{S}^{{}^{-1,\starT}}\starT\mathscr{S}&=-\mathscr{S}^{{}^{-1,\starT}}\starT\partial_j\mathscr{S}\\&=\frac{ig}{2\hbar}\phi_j^2+{g}\Delta^R_{jk}\phi_j\ \partial_k\mathscr{S}^{{}^{-1,\starT}}\starT\mathscr{S}-\frac{i\hbar g}{2}\Delta^R_{jk}\Delta^R_{jl}\ \partial^2_{kl}\mathscr{S}^{{}^{-1,\starT}}\starT\mathscr{S},\end{split}
\end{equation}
where to get the second line we set $V=\frac{1}{3!}\sum_a\phi_a^3$ (so that $\partial_j\mathscr{S}=-\frac{ig}{2\hbar}\phi_j^2 \mathscr{S}$), wrote out the $\starT$ product as an expansion in $\hbar$, and grouped the terms by the number of derivatives acting on $\phi_j^2$. Note that in the resulting expression, there is implicit summation over the repeated $k,l$ indices but not over $j$.

Plug this into \eqref{eq: derivation 2}, noting that $\Delta^R_{ab}=0$ if $a<b$, to obtain,
\begin{equation}
    R_{\mathcal{T},gV}(\phi_3)=\phi_3+\frac{g}{2}\Delta^R_{3j}\phi_j^2-ig\hbar\Delta^R_{32}\Delta^R_{21}\phi_2\ \partial_1\mathscr{S}^{{}^{-1,\starT}}\starT\mathscr{S}-\frac{g\hbar^2}{2}\Delta^{R}_{32}(\Delta^R_{21})^2\ \partial_1^2\mathscr{S}^{{}^{-1,\starT}}\starT\mathscr{S},
\end{equation}
where now we sum over the repeated $j$ index. The third term on the RHS can be further simplified by setting $j=1$ in \eqref{eq: derivation 1} to find $\partial_1\mathscr{S}^{{}^{-1,\starT}}\starT\mathscr{S}=\frac{ig}{2\hbar}\phi_1^2$. To simplify the final term, apply the Leibniz rule to $\partial^2_{1}(\mathscr{S}^{{}^{-1,\starT}}\starT\mathscr{S})=0$ to obtain,
\begin{equation}\label{eq: 12082}
\partial^2_{1}\mathscr{S}^{{}^{-1,\starT}}\starT\mathscr{S}=-2\partial_{1}\mathscr{S}^{{}^{-1,\starT}}\starT\partial_{1}\mathscr{S}-\mathscr{S}^{{}^{-1,\starT}}\starT\partial^2_{1}\mathscr{S}.
\end{equation}
The first term on the RHS is simplified by noting again that $\partial_j\mathscr{S}=-\frac{ig}{2\hbar}\phi_j^2 \mathscr{S}$ for our choice of $V$, and therefore, $\partial_{1}\mathscr{S}^{{}^{-1,\starT}}\starT\partial_{1}\mathscr{S}=-\frac{ig}{2\hbar}\phi_1^2\partial_{1}\mathscr{S}^{{}^{-1,\starT}}\starT \mathscr{S}$, where we could pull the $\phi_1^2$ outside of the star product because $\Delta^R_{1a}=0$ for all $a$. To further simplify this, set $j=1$ in \eqref{eq: derivation 1}.

 The second term on the RHS of \eqref{eq: 12082} is simplified by noting that our choice of $V$ gives $\partial_1^2\mathscr{S}=-\frac{ig}{\hbar}(\phi_1-\frac{ig}{4\hbar}\phi_1^4) \mathscr{S}$ and therefore $\mathscr{S}^{{}^{-1,\starT}}\starT\partial^2_{1}\mathscr{S}=-\frac{ig}{\hbar}(\phi_1-\frac{ig}{4\hbar}\phi_1^4).$
Thus altogether we obtain,
\begin{equation}\label{eq from example RT}
\begin{split}
R_{\mathcal{T},gV}(\phi_3)=\phi_3+\frac{g}{2} \, \Delta^R_{3j}\phi_{j}^2 +\frac{g^2}{2}\Delta^R_{32}\Delta^R_{21}\big( \phi_2\phi_1^2-i\hbar\Delta^R_{21}\phi_1\big)+\frac{g^3}{8}(\Delta^R_{32})(\Delta^R_{21})^2\phi_1^4.
\end{split}\end{equation}
Applying the time ordering operator $\mathcal{T}$ to both sides of \eqref{eq from example RT} gives,
\begin{equation}
    \begin{split}
    \mathcal{T}R_{\mathcal{T},gV}(\phi_3)
   &=\phi_3+\frac{g}{2} \, \sum_{j=1,2}\Delta^R_{3j}(\phi_{j}^2+\hbar \Delta^F_{jj}) +\frac{g^2}{2}\Delta^R_{32}\Delta^R_{21}\big( \phi_2\phi_1^2+2\hbar\Delta^F_{21}\phi_1+\hbar\Delta^F_{11}\phi_2-i\hbar\Delta^R_{21}\phi_1\big)\\
   &\ \ \ +\frac{g^3}{8}(\Delta^R_{32})(\Delta^R_{21})^2(\phi_1^4+6\hbar\Delta^F_{11}\phi_1^2+3\hbar^2(\Delta^F_{11})^2).
    \end{split}
\end{equation}
Finally, set $\phi=0$ to extract the constant term and obtain our answer,
\begin{equation}
    \langle \hatphi_3^H \rangle=\frac{g\hbar}{2} \, \Delta^R_{32} \Delta^F_{22}+\frac{g\hbar}{2} \, \Delta^R_{31} \Delta^F_{11}+\frac{3g^3\hbar^2}{8}\Delta^R_{32}(\Delta^R_{21})^2(\Delta^F_{11})^2.
\end{equation}
\end{example}

\section{Diagrammatic expansions}\label{sec: diagramatic}
As we have seen, computations in the operator formalism can be translated into equivalent computations in the functional algebraic formalism of pAQFT. In Section \ref{sec: examples} we showed how these can be computed in practice by evaluating M\o ller operators. A different, powerful method for computations is through a diagrammatic expansion. Diagrammatic expansions for $\mathcal{T}(F)$, $R_{\mathcal T,gV}(F)$ and $R_{H,gV}(F)$ have been derived in \cite{Hawkins:2016ddq}. Independently, diagrammatic expansions for in-in and in-out correlators have been derived using the operator approach in \cite{Albertini:2024srq}. In this section we present these expansions and show their equivalence, providing an alternative proof to Proposition \ref{proposition: equivlence} in Corollary~\ref{equivalence again}.

Throughout this section we assume that $V=\sum_{z\in\calC}c_z\phi_z^{v}$ is homogeneous of a single degree $v$, the \emph{order} of $V$; the constants $c_z$ enter through the vertex factors. The general case is easy to obtain by linearity.
\subsection{Operator diagrammatic expansions}
We begin with some definitions.

\begin{definition}[Half-edge graph, adapted from \cite{Yeats}]\label{half edge graph}
A half-edge graph $G$ is a set of half edges along with a set $V(G)$ of disjoint subsets of half edges known as vertices which partition the
set of half edges, and a set $E(G)$ of disjoint pairs of half edges known as internal edges. The pairs may be ordered or unordered, corresponding to directed or undirected edges.
Those half edges which are not in any internal edge are known as external edges (these are always undirected). 
\end{definition}

\begin{definition}[Half-edge factor]\label{def H}
    Given a half-edge graph $G$, its half-edge factor $H(G)$ is the number of ways that the half edges could be paired into internal edges, keeping $V(G)$ and the number and direction of edges between vertices fixed.
\end{definition}
\begin{remark}
    We will only consider graphs up to isomorphism. These are represented as diagrams in which the vertices and edges are unlabelled. We may decorate these diagrams by assigning labels to some or all nodes. We call the decorated nodes external vertices and the undecorated vertices internal vertices.
\end{remark}

\begin{definition}[Vertex permutation]\label{vertex permuations}
 Given a diagram $D$, assign a label to each internal node. A bijection from this label set to itself is called a vertex permutation. We write $\VP(D)$ to denote the subgroup of vertex permutations that leave the labelled diagram invariant.
   \end{definition}

\begin{definition}
    [Operator Feynman rules]\label{def Feynman} Given a half-edge graph $G$, the operator Feynman rules assign a value to it as follows:
\begin{itemize}
\item each internal vertex is assigned a label $z\in\calC$ and contributes the factor $+\frac{ig}{\hbar}c_z$.
	\item A diagram with internal vertices labelled $z_1,\ldots,z_n$ carries $\sum_{z_1,\ldots,z_n\in\calC}$, with the factors $\frac{ig}{\hbar}c_{z_1},\ldots,\frac{ig}{\hbar}c_{z_n}$ inside the sum,
\item each directed edge $a\rightarrow b$ gives a factor of $-i\hbar\Delta^R_{ba}$,
\item  each undirected edge $a\mathrel{\mbox{---}}b$  gives a factor of $\hbar\Delta^F_{ab}$,
\item each external edge at an (internal or external) vertex $z$ gives a field operator $\hatphi_z$; the product of these factors is causally ordered,
\item multiply by the half-edge factor $H(G)$ (as given in definition \ref{def H}),
\item divide by $\left|\VP(G)\right|$ (as given in definition \ref{vertex permuations}).\footnote{In \cite{Albertini:2024srq}, $\abs{\VP(G)}$ was denoted $\abs{\Aut(G)}$.}
\end{itemize}
\end{definition}

\subsection{Algebraic diagrammatic expansions}
While in the operator approach graphs are used to represent operators and expectation values, in the algebraic approach graphs are used to represent multidifferential operators acting on functionals. In this context, it is useful to define the notion of graph as follows.

\begin{definition}[Graph, automorphism]
A \emph{directed graph} consists of two finite sets, $V$ (the set of vertices) and $E$ (the set of edges) and two functions $\source,\target:E\to V$. Graphically, the vertices are represented as points, and $e\in E$ is represented as an arrow from $\source(e)$ to $\target(e)$. An \emph{automorphism} of such a graph is a pair of bijections, $a_V:V\to V$ and $a_E:E\to E$ such that $\source\circ a_E=a_V\circ\source$ and $\target\circ a_E=a_V\circ\target$.

An \emph{undirected graph} consists of a directed graph and a function, $\inv:E\to E$ with no fixed points, such that for all $e\in E$, $\inv\circ\inv(e)=e$ and $\source\circ\inv = \target$. An undirected edge is an (unordered) pair $\{e,\inv(e)\}$ and is represented graphically as a curve connecting $\source(e)$ and $\target(e)$. An \emph{automorphism} of an undirected graph is an automorphism of the underlying graph such that $\inv\circ a_E=a_E\circ\inv$.

A graph with multiple types of edges consists of multiple graphs (directed or undirected) with the same vertex set.

For $k\in\NN$, $\Gcal(k)$ denotes the set of isomorphism classes of graphs with \emph{external} vertices labelled $1,\dots,k$, and possibly unlabelled \emph{internal} vertices, directed edges, and undirected edges. For $\gamma\in\Gcal(k)$, $\Aut\gamma$ denotes the group of automorphisms of $\gamma$ that leave the external vertices fixed.
\end{definition}

The translation from graph to differential operator is given by the algebraic Feynman rules.

\begin{definition}[Algebraic Feynman rules]\label{algebraic Feynman rules}
For  $\gamma\in\Gcal(k)$  and an interaction polynomial $V\in\CC[\calC]$,  $\vec\gamma:\CC[\calC]^{\times k}\to\CC[\calC]$ is a multidifferential operator, where $\vec\gamma(F_1,\dots,F_k)$ is obtained by,
\begin{itemize}
    \item  assign a dummy index to each end of an edge, 
     \item the edge $i\rightarrow j$ is evaluated as $-i\hbar \Delta^R_{ji}$, so that arrows point from the past to the future, as in Definition~\ref{def Feynman},
    \item the edge $i\mathrel{\mbox{---}}j$ is evaluated  as $\hbar\Delta^F_{ij}$,
    \item an internal vertex where the edge ends labelled $i_1,\ldots,i_r$ meet is evaluated as $\frac{ig}{\hbar}\partial^r_{i_1,\ldots,i_r}V$,
    \item the $n^\ith$  external vertex, where $i_1,\ldots,i_r$ meet is evaluated as $\partial^r_{i_1,\ldots,i_r}F_n$,
    \item the dummy indices are summed over,
    \item divide by $\lvert\Aut\gamma\rvert$.
\end{itemize}
\end{definition}

\begin{remark}    The Algebraic Feynman rules given in definition \ref{algebraic Feynman rules} are adapted from \cite[Defs.~6.3 and 6.37]{Hawkins:2016ddq}. Factors of $i,\hbar$, $g$, and $\abs{\Aut\gamma}$ were not included in the the Feynman rules there but appeared explicitly as factors multiplying the diagrams in the sum. Here, we include these factors in the Feynman rules. \cite{Hawkins:2016ddq} use $\mathscr S=e^{\frac{i\lambda}{\hbar}V}$, so $\lambda=-g$ in our conventions.  In \cite[Def.~6.3]{Hawkins:2016ddq} an edge from $y$ to $x$ represents $\Delta^A_{S_0}(x,y)$, so it is nonvanishing only when it points from the future to the past. Here a directed edge $i\to j$ represents $-i\hbar\Delta^R_{ji}=-i\hbar\Delta^A_{ij}$ and points from the past to the future, as in the operator rules and in \cite{Albertini:2024srq}.
\end{remark}

\begin{example}\label{example: product}
If $\gamma\in\Gcal(k)$ is the graph with no internal vertices and no edges, then
\begin{equation}
\vec\gamma(F_1,\dots,F_k) = F_1\dots F_k .
\end{equation}
\end{example}

\subsection{Comparison}
The following lemmas will be useful in the comparison of the two diagrammatic approaches.
\begin{lemma}\label{lemma: H}
    Let $G$ be a half-edge graph with $n$ vertices of valency $v_1,\ldots,v_n$ and no closed directed cycles. Then the half-edge factor $H(G)$ is equal to,
   \begin{equation}
       H(G)=\frac{\prod_{i} v_i!}{\prod_{i<j} k_{ij}!\cdot \prod_{i,j} k_{i\rightarrow j}!\cdot\prod_{i}2^{k_i} k_i!h_i!},
   \end{equation}
where $k_{ij}$  {\normalfont(}$k_{i\rightarrow j}!${\normalfont)} is the number of undirected {\normalfont(}directed{\normalfont)} edges connecting the $i^\ith$ and $j^\ith$ vertices, $k_i$ is the number of undirected loops from the $i^\ith$ vertex to itself and $h_i$ is the number of  unpaired external half-edges at the $i^\ith$ vertex.
\end{lemma}
\begin{proof}
    First, consider connecting the half-edges to form undirected edges from each vertex to itself. There are, \begin{equation}\label{eq: H proof}
      \prod_{i=1}^n  \frac{1}{k_i!}\binom{v_i}{2}\binom{v_i-2}{2}\cdots\binom{v_i-2k_i+2}{2}=\prod_{i=1}^n \frac{v_i!}{(v_i-2k_i)!k_i!2^{k_i}},
    \end{equation} ways to do this. Now consider connecting a pair of vertices $a,b$ with undirected edges. This gives a factor of,
    \begin{equation}
      k_{ab}! \binom{v_a-2k_a}{k_{ab}} \binom{v_b-2k_b}{k_{ab}} =\frac{(v_a-2k_a)!(v_b-2k_b)!}{k_{ab}!(v_a-2k_a-k_{ab})!(v_b-2k_b-k_{ab})!}.
    \end{equation}
Multiplying with \eqref{eq: H proof} ``replaces'' the factor $(v_a-2k_a)!(v_b-2k_b)!$ in the denominator of \eqref{eq: H proof} with $k_{ab}!(v_a-2k_a-k_{ab})!(v_b-2k_b-k_{ab})!$. As we continue to add edges between vertices, we will similarly gain the remaining factors of $k_{ij}!$ and $k_{i\rightarrow j}!$ in the denominator, while factors of the form $(v_a-x_a)!$ will continue to be ``replaced'' by factors of the same form with larger and larger $x_a$ until they are equal $h_i!$ for all $i$. This gives the desired form.
\end{proof}

\begin{lemma}\label{lemma: aut}
       Let $\gamma$ be a graph with vertex set $V$. The number of automorphisms is
       \begin{equation}
   \left|\Aut\gamma\right| =  \bigl|\VP(\gamma)\bigr|\prod_{i< j\in V} k_{ij}! \prod_{i,j\in V} k_{i\rightarrow j}!\prod_{i\in V}2^{k_i} k_i!,
       \end{equation}
       where $\left|\VP(\gamma)\right|$ is as defined in  \ref{vertex permuations}, $k_{ij}$  {\normalfont(}$k_{i\rightarrow j}${\normalfont)} is the number of undirected {\normalfont(}directed{\normalfont)} edges connecting the $i^\ith$ and $j^\ith$ vertices and $k_i$ is the number of undirected loops from the $i^\ith$ vertex to itself. 
\end{lemma}
\begin{proof}
There is a homomorphism from $\Aut\gamma$ to the group of permutations of $V$, given by $a\mapsto a_V$. By construction, $\VP\gamma$ is the image of this homomorphism. Let $\Aut_0\gamma$ be the kernel, so that $\VP\gamma = \Aut\gamma/\Aut_0\gamma$. This gives, $\abs{\Aut\gamma} = \abs{\VP\gamma}\abs{\Aut_0\gamma}$. $\Aut_0\gamma \subseteq \Aut\gamma$ is the subgroup of automorphisms that leave all vertices fixed.

For $i,j\in V$, there is a subgroup of $\Aut_0\gamma$ consisting of the automorphisms that fix all edges except directed edges from $i$ to $j$. This subgroup consists of all permutations of the set of directed edges from $i$ to $j$, so its order is $k_{i\to j}!$.

Similarly, for each unordered pair $\{i,j\}\subseteq V$ with $i\neq j$, there is a subgroup consisting of the automorphisms that fix all edges except undirected edges between $i$ and $j$. Its order is $k_{ij}!$.

Finally, for $i\in V$, there is a subgroup of the automorphisms that fix all edges except undirected edges between $i$ and itself. These automorphisms permute these edges from $i$ to itself, but at the same time it can flip edges. This subgroup is isomorphic to a semidirect product of the symmetric group on a set of $k_i$ edges with $\ZZ_2^{k_i}$, so its order is $2^{k_i}k_i!$.

$\Aut_0\gamma$ is the product of all of these subgroups, so the order $\abs{\Aut\gamma}$ is the product of all of these factors.
\end{proof}

\subsection{The time ordering operator and in-out correlators}
This section gives the diagrammatics of equation \eqref{proposition: in out}.
\begin{lemma}[Diagrammatic expansion of in-out correlators, from \cite{Albertini:2024srq}]\label{lemma: diagram in out}
Given $x_i\not\prec x_j$ for every $i<j$, some positive integers $m_1,\ldots, m_k$ and an interaction polynomial $V$,  the expectation value, $$_{\out}\big\langle 0\big|(\phi_{x_1}^H)^{m_1}\ldots (\phi_{x_k}^H)^{m_k}\big|0\big\rangle_{\IN},$$ is equal to the sum of undirected half-edge graphs without external edges and with $k$ external vertices labelled $x_1,\ldots,x_k$, each with valency $m_1,\ldots, m_k$, respectively, and any number of unlabelled internal vertices,
  each with valency equal to the order of the interaction polynomial $V$, where each graph $\gamma$ is assigned a value via the operator Feynman rules and multiplied by $(-1)^{v(\gamma)}$, $v(\gamma)$ being the number of internal vertices.
\end{lemma}
\begin{proof}
By ${}_\out\langle0|=\langle0|\hat S$ and Lemma~\ref{lemma: causal factorisation}(ii), the expectation value equals $\langle0|C[\hat S\,\hatphi_{x_1}^{m_1}\cdots\hatphi_{x_k}^{m_k}]|0\rangle$. Expand $\hat S=C[\prod_ze^{-\frac{ig}{\hbar}V_z(\hatphi_z)}]$ and apply Wick's theorem for causally ordered products, whose contractions are $\hbar\Delta^F$. The half-edge factor counts the pairings, and the passage from labelled to unlabelled internal vertices produces $1/\lvert\VP(G)\rvert$. Each internal vertex carries the factor $-\frac{ig}{\hbar}c_z$ from $\hat S$, which is $-1$ times the vertex factor of Definition~\ref{def Feynman}; this gives $(-1)^{v(G)}$.
\end{proof}

\begin{lemma}[Diagrammatic expansion of $\mathcal{T}(\mathscr{S}\phi_{x_1}^{m_1}\cdots\phi_{x_k}^{m_k})(0)$]\label{lemma: diagram T}Given $x_i\not\prec x_j$ for every $i<j$, some positive integers $m_1,\ldots, m_k$ and an interaction polynomial $V$,  $$\mathcal{T}(\mathscr{S}\phi_{x_1}^{m_1}\cdots\phi_{x_k}^{m_k})(0),$$ is equal to the sum of undirected graphs with $k$ external vertices labelled $x_1,\ldots,x_k$, each with valency $m_1,\ldots, m_k$, respectively, and
any number of unlabelled internal vertices,
  each with valency equal to the order of the interaction polynomial $V$, where each graph is assigned a value via the Algebraic Feynman Rules  and multiplied by $(-1)^{v(\gamma)}$.
\end{lemma}

To prove \ref{lemma: diagram T}, we will need the following lemma. This is Lemma~6.13 of \cite{Hawkins:2016ddq} rewritten to take into account the powers of $i$, $\hbar$, $g$, and $\abs{\Aut\gamma}$ in Definition~\ref{algebraic Feynman rules}.
\begin{lemma}
\label{Composition Lemma}
If $\alpha$ and $\gamma$ are graphs, and $j$ is one of the external vertices of $\gamma$, then the composition of $\vec\alpha$ into the $j$'th argument of $\vec\gamma$ is 
\begin{equation}
\vec\gamma\circ_j\vec\alpha = \sum_\beta d_\beta \vec\beta
\end{equation}
where the sum is over graphs $\beta$ that can be obtained as an extension of $\gamma$ by $\alpha$ at $j$ --- that is, $\beta$ is obtained by removing the vertex $j$ and attaching the adjacent edges to vertices of $\alpha$ --- and $d_\beta$ is the number of different ways of obtaining $\beta$ as such an extension.
\end{lemma}
In \cite{Hawkins:2016ddq} this lemma is stated for graphs with directed edges only. Its proof uses only that the edge kernels are constant (and the product rule), so it applies verbatim to graphs with undirected edges, as used below.
\begin{lemma}\label{T lemma}
\cite[Lem.~6.39]{Hawkins:2016ddq}
\[
\Tcal = \sum_{\gamma\in\Gcal_6(1)}  \vec\gamma ,
\]
where a graph is in $\Gcal_6(1)\subset\Gcal(1)$ if it has only 1 vertex and undirected edges.
\end{lemma}
\begin{corollary}
\label{lemma T graph}
If $F_1,\dots,F_k\in \CC[\calC]$ and $\gamma\in\Gcal(k)$, then 
\[
\mathcal{T}(\vec\gamma(F_1,\dots,F_k)) = \sum_\alpha \vec\alpha(F_1,\dots,F_k)
\]
where the sum is over graphs that are obtained from $\gamma$ by adding undirected Feynman edges, including undirected loops, in all possible ways.  
\end{corollary}
\begin{proof}
This follows directly from Lemmas \ref{Composition Lemma} and \ref{T lemma}.
\end{proof}
\begin{lemma}[Diagrammatic product rule]\label{lemma: product rule} Given some $F_1,\dots,F_k\in\CC[\calC]$, and a graph $\gamma\in\Gcal(1)$, 
\[
\vec\gamma(F_1\cdots F_k)=\sum_{\alpha}\vec\alpha(F_1,\ldots,F_k)
\]
where the sum is over all graphs $\alpha\in\Gcal(k)$ that can be obtained from $\gamma$ by replacing the single external vertex with $k$ external vertices, keeping the valency of the internal vertices, the total number of edges and their direction as they were in $\gamma$.
   \end{lemma}
\begin{proof}
This is immediate from Lemma~\ref{Composition Lemma} and Example~\ref{example: product}.
\end{proof}

\begin{proof}[Proof of Lemma \ref{lemma: diagram T}]
Since $\mathcal{T}$ is linear write,
\begin{equation}
    \mathcal{T}(\mathscr{S}\phi_{x_1}^{m_1}\cdots\phi_{x_k}^{m_k})=\sum_{n=0}^{\infty}\mathcal{T}\left(\big(\tfrac{-ig}{\hbar}\big)^n\tfrac{1}{n!}V^n\phi_{x_1}^{m_1}\cdots\phi_{x_k}^{m_k}\right).
\end{equation}
By the Algebraic Feynman rules (definition \ref{algebraic Feynman rules}), $\big(\frac{-ig}{\hbar}\big)^n\frac{1}{n!}V^n\phi_{x_1}^{m_1}\cdots\phi_{x_k}^{m_k}$ is given by the graph that has $n$ internal vertices, $k$ external vertices labelled $x_1,\cdots,x_k$ and no edges. By Corollary \ref{lemma T graph}, $\mathcal{T}(\big(\frac{-ig}{\hbar}\big)^n\frac{1}{n!}V^n\phi_{x_1}^{m_1}\cdots\phi_{x_k}^{m_k})$ is given by the sum of graphs with $n$ internal vertices, $k$ external vertices  $x_1,\cdots,x_k$ and any number of undirected edges. Setting $\phi=0$ sets to zero all graphs in which the valency of any of the internal vertices is less than the order of $V$ or in which the valency of the vertex $x_i$ is less than $m_i$. This completes the proof.
\end{proof}

\begin{corollary}\label{equivalence again}
    Given $x_i\not\prec x_j$ for every $i<j$, the diagrammatic expansion of $$_{\out}\big\langle 0\big|(\phi_{x_1}^H)^{m_1}\ldots (\phi_{x_k}^H)^{m_k}\big|0\big\rangle_{\IN}$$ in the operator approach and of $\mathcal{T}(\mathscr{S}\phi_{x_1}^{m_1}\cdots\phi_{x_k}^{m_k})(0)$ in the algebraic approach are identical (hence Proposition~\ref{proposition: equivlence} is true).
\end{corollary}
\begin{proof}
     By comparison of Lemmas \ref{lemma: diagram in out} and \ref{lemma: diagram T}, the diagrams in the two expansions are identical. By comparison of the operator and algebraic Feynman rules (definitions \ref{def Feynman} and \ref{algebraic Feynman rules}) we find that both assign $\hbar\Delta^F$ to undirected edges, they end up with the same coefficient $\frac{ig}{\hbar}c_z$ at internal vertices, the same sign $(-1)^{v(D)}$ and the same summation over indices. In the algebraic approach, each diagram $D$ with $n$ internal vertices additionally encodes within it the constant, $\frac{(v!)^n\prod_i m_i!}{\abs{\Aut(D)} }$, where $v$ is the order of $V$ (the factor $m_i!$ comes from the external vertex $x_i$ after $m_i$ derivatives have been applied to $\phi_{x_i}^{m_i}$) and $\abs{\Aut (D)}$ is the order of the automorphism group that leaves external vertices fixed. In the operator approach, each diagram encodes within it the constant $\frac{H(D)}{\abs{\VP(D)}}$, where $H$ is the half-edge factor and $\VP$ is the number of vertex permutations that leaves external vertices fixed. That $\frac{(v!)^n\prod_i m_i!}{\abs{\Aut(D)}}=\frac{H(D)}{\abs{\VP(D)}}$ for any diagram $D$ in this expansion follows from Lemmas \ref{lemma: H} (with $h_i=0$) and~\ref{lemma: aut}.
\end{proof}

We finish this section with a concise formula for the normalised in-out correlator in terms of algebraic Feynman diagrams. 
\begin{proposition}
Let $x_i\not\prec x_j\in\calC$ for every $i<j$ and functionals $F_1,\dots,F_n\in\CC[\calC]$ such that $\supp F_k=\{x_k\}$.  The normalised out-in-expectation value of the product of the corresponding Heisenberg operators is
\begin{align}
{}_{\out}\langle0\rvert F_1(\hat\phi^H) \dots F_n(\hat\phi^H)\lvert0\rangle^{\mathrm{norm}}_{\IN} 
&= \sum_{\gamma} (-1)^{v(\gamma)} \vec\gamma(F_1,\dots,F_n)(0) \label{out-in expectation}
\end{align}
where the sum is over undirected $\gamma\in\Gcal(n)$ such that every connected component contains an external vertex.
\end{proposition}
\begin{proof}
By Proposition~\ref{proposition: equivlence},
\[
{}_{\out}\langle0\rvert F_1(\hat\phi^H) \dots F_n(\hat\phi^H)\lvert0\rangle_{\IN} = \Tcal(\mathscr S F_1\dots F_n)(0) ,
\]
so
\[
{}_{\out}\langle0\rvert F_1(\hat\phi^H) \dots F_n(\hat\phi^H)\lvert0\rangle^{\mathrm{norm}}_{\IN}
= \frac{\Tcal(\mathscr S F_1\dots F_n)}{\Tcal\mathscr S}(0) .
\]

Now we can start writing diagrammatic formulae. The exponential for the S-matrix is
\[
\mathscr S = e^{-i\frac g\hbar V} 
= \sum_{\gamma} (-1)^{v(\gamma)}\vec\gamma ,
\]
where the sum is over $\gamma\in\Gcal(0)$ with no edges. Multiplying with some $F\in\CC[\calC]$, this trivially gives
\[
\mathscr S F = e^{-i\frac g\hbar V} F 
= \sum_{\gamma} (-1)^{v(\gamma)}\vec\gamma(F) ,
\]
where the sum is over $\gamma\in\Gcal(1)$ with no edges. Now, composing with $\Tcal$, Lemma~\ref{T lemma} gives
\[
\Tcal(\mathscr SF) = \sum_{\gamma} (-1)^{v(\gamma)}\vec\gamma(F) ,
\]
where the sum is over all undirected $\gamma\in\Gcal(1)$. Any such graph decomposes into the connected component of the external vertex and an undirected graph in $\Gcal(0)$. In particular, this shows that $\Tcal(\mathscr S)$ is given by the sum over undirected graphs in $\Gcal(0)$ and 
\[
\frac{\Tcal(\mathscr SF)}{\Tcal(\mathscr S)} = \sum_{\gamma} (-1)^{v(\gamma)} \vec\gamma(F) ,
\]
where the sum is over \emph{connected} undirected graphs $\gamma\in\Gcal(1)$. 

Now set $F=F_1\dots F_n$. By Lemma~\ref{lemma: product rule} (the diagrammatic product rule),
\[
\frac{\Tcal(\mathscr SF_1\dots F_n)}{\Tcal(\mathscr S)} = \sum_{\gamma} (-1)^{v(\gamma)}\vec\gamma(F_1,\dots,F_n) ,
\]
where the sum is over undirected $\gamma\in\Gcal(n)$ such that every connected component contains an external vertex. 
Evaluating at $\phi=0$ gives the desired result.
\end{proof}

\subsection{$R_{\mathcal{T},gV}$ and operator expansion}
In this section we give a diagrammatic proof to the following lemma.

\begin{lemma}\label{lemma: RT and operators}
Given $x_i\not\prec x_j\in\calC$ for every $i<j$, some positive integers $m_1,\ldots, m_k$ and an interaction polynomial $V$,
\begin{equation}
    (\hatphi_{x_1}^H)^{m_1}\ldots (\hatphi_{x_k}^H)^{m_k}=\QT R_{\mathcal{T},gV}(\phi_{x_1}^{m_1}\ldots \phi_{x_k}^{m_k}).
\end{equation}
\end{lemma}

It will be helpful to define the following class of diagrams.

\begin{definition}[Sink graphs]\label{def: sink}
A sink graph is a directed graph in which all external vertices are sinks, every internal vertex has at least one outgoing edge and there are no closed directed cycles. A sink half-edge graph is as above, with the addition that (undirected) external edges are allowed.
\end{definition}
\begin{remark}
    In a sink graph, for every internal vertex there is at least one directed path connecting it to some external vertex.
\end{remark}

The diagrammatic expansion of the LHS is given by the following lemma.

\begin{lemma}(Diagrammatic expansion of causally ordered operator product.)\label{lemma: diagrams operator expansion}
    Given $x_i\not\prec x_j\in\calC$ for every $i<j$, some positive integers $m_1,\ldots, m_k$ and an interaction polynomial $V$,  the operator product 
  $(\phi_{x_1}^H)^{m_1}\ldots (\phi_{x_k}^H)^{m_k}$ is equal to the sum of sink half-edge graphs, with $k$ external vertices of valencies $m_1,\cdots,m_k$ and any number of internal vertices of valency equal to the order of $V$, where each graph is assigned a value via the operator Feynman rules.
\end{lemma}
Lemma \ref{lemma: diagrams operator expansion} is derivative from the work of \cite{Albertini:2024srq}, although it has not been stated explicitly in that work. For completeness, we give its derivation in appendix \ref{appendix: sec diagrams}.

Now consider the RHS.

\begin{lemma}[Diagrammatic representation $R_{\mathcal{T},gV}(\phi_{x_1}^{m_1}\ldots \phi_{x_k}^{m_k})$]\label{lemma:2008}
$R_{\mathcal{T},gV}(\phi_{x_1}^{m_1}\ldots \phi_{x_k}^{m_k})$ is equal to the sum of operators $\vec{\gamma}(\phi_{x_1}^{m_1},\ldots, \phi_{x_k}^{m_k})$ defined by sink graphs $\gamma\in\Gcal(k)$ via the algebraic Feynman rules.
   \end{lemma}

   To prove Lemma \ref{lemma:2008} we will need the following.
\begin{lemma}[Diagrammatic representation $R_{\mathcal{T},gV}(F)$, from \cite{Hawkins:2016ddq}]\label{lemma: 2108}
\begin{equation}
R_{\mathcal{T},gV}(F) = \sum_\gamma \vec{\gamma}(F)
\end{equation} 
where the sum is over sink graphs $\gamma\in\Gcal(1)$.
\end{lemma}

  \begin{proof}[Proof to Lemma \ref{lemma:2008}]
  Follows immediately from combining Lemmas \ref {lemma: product rule} and \ref{lemma: 2108}.
  \end{proof}
\begin{example} We give an illustration of the product rule (Lemma \ref{lemma: product rule}) applied to the diagrammatic expansion of $R_{\mathcal{T},gV}$ (as given in Lemma \ref{lemma: 2108}). Let $F(\phi)=\phi_x\phi_y$. Then, 
\begin{equation}\label{eq example: moller non local}
\begin{split}
R_{\mathcal{T},gV}(F) =& \quad
 \begin{tikzpicture}[baseline=(current bounding box.center), decoration={markings,
    mark= at position 0.6 with {\arrow{Straight Barb[ length=1mm, width=1mm]}}}
] 
  \tikzstyle{every node}=[circle, draw, fill=black,inner sep=0pt, minimum width=3pt]
  \node (n3) at (9,0) [label= right: $F$]{};
\end{tikzpicture}
\quad +\;
\begin{tikzpicture}
[baseline=(current bounding box.center),decoration={markings, 
    mark= at position 0.8 with {\arrow{Straight Barb[ length=1mm, width=1mm]}}}
] 
  \tikzstyle{every node}=[circle, draw, fill=black,inner sep=0pt, minimum width=3pt]
  \node (n3) at (9,0.5) [label= right: $F$]{};
  \node(n2) at (9,0)  [label= right: ]{};
    \draw [postaction={decorate}](n2) -- (n3)  {} ;
\end{tikzpicture}
\quad +\;
\begin{tikzpicture}
[baseline=(current bounding box.center),decoration={markings, 
    mark= at position 0.6 with {\arrow{Straight Barb[ length=1mm, width=1mm]}}}
] 
  \tikzstyle{every node}=[circle, draw, fill=black,inner sep=0pt, minimum width=3pt]
  \node[align=left] (n2) at (9,0.5)  [label= right: $F$]{};
   \node[align=left]  (n1) at (9,0)  [label= right: ]{};
    \draw [postaction={decorate}](n2) -- (n3)  {} ;
    \draw [postaction={decorate}](n1) to[bend left=60] (n2)  {} ;
     \draw [postaction={decorate}](n1) to[bend right=60] (n2)  {} ;
      ;
\end{tikzpicture}+\cdots
\\
=& \quad
 \begin{tikzpicture}[baseline=(current bounding box.center), decoration={markings,
    mark= at position 0.6 with {\arrow{Straight Barb[ length=1mm, width=1mm]}}}
] 
  \tikzstyle{every node}=[circle, draw, fill=black,inner sep=0pt, minimum width=3pt]
  \node (n3) at (9,0) [label= right: $\phi_x$]{};
  \node (n3) at (9.75,0) [label= right: $\phi_y$]{};
\end{tikzpicture}
\quad+ \; \begin{tikzpicture}
[baseline=(current bounding box.center),decoration={markings, 
    mark= at position 0.8 with {\arrow{Straight Barb[ length=1mm, width=1mm]}}}
] 
  \tikzstyle{every node}=[circle, draw, fill=black,inner sep=0pt, minimum width=3pt]
  \node (n3) at (9,0.5) [label= right: $\phi_x$]{};
  \node (n4) at (9.75,0.5) [label= right: $\phi_y$]{};
  \node(n2) at (9,0)  [label= right: ]{};
    \draw [postaction={decorate}](n2) -- (n3)  {} ;
\end{tikzpicture}\quad+\begin{tikzpicture}
[baseline=(current bounding box.center),decoration={markings, 
    mark= at position 0.8 with {\arrow{Straight Barb[ length=1mm, width=1mm]}}}
] 
  \tikzstyle{every node}=[circle, draw, fill=black,inner sep=0pt, minimum width=3pt]
  \node (n3) at (9,0.5) [label= right: $\phi_y$]{};
  \node (n4) at (9.75,0.5) [label= right: $\phi_x$]{};
  \node(n2) at (9,0)  [label= right: ]{};
    \draw [postaction={decorate}](n2) -- (n3)  {} ;
\end{tikzpicture}\quad+\;
\begin{tikzpicture}
[baseline=(current bounding box.center),decoration={markings, 
    mark= at position 0.8 with {\arrow{Straight Barb[ length=1mm, width=1mm]}}}
] 
  \tikzstyle{every node}=[circle, draw, fill=black,inner sep=0pt, minimum width=3pt]
  \node (n3) at (9,0.5) [label= right: $\phi_x$]{};
  \node (n4) at (9.75,0.5) [label= right: $\phi_y$]{};
  \node(n2) at (9.37,0)  [label= right: ]{};
    \draw [postaction={decorate}](n2) -- (n3)  {} ;
     \draw [postaction={decorate}](n2) -- (n4)  {} ;
\end{tikzpicture}+\begin{tikzpicture}
[baseline=(current bounding box.center),decoration={markings, 
    mark= at position 0.6 with {\arrow{Straight Barb[ length=1mm, width=1mm]}}}
] 
  \tikzstyle{every node}=[circle, draw, fill=black,inner sep=0pt, minimum width=3pt]
  \node[align=left] (n2) at (9,0.5)  [label= right: $\phi_x$]{};
   \node[align=left]  (n1) at (9,0)  [label= right: ]{};
    \draw [postaction={decorate}](n2) -- (n3)  {} ;
    \draw [postaction={decorate}](n1) to[bend left=60] (n2)  {} ;
     \draw [postaction={decorate}](n1) to[bend right=60] (n2)  {} ;
      ; \node (n4) at (9.75,0.5) [label= right: $\phi_y$]{};
\end{tikzpicture}
+\begin{tikzpicture}
[baseline=(current bounding box.center),decoration={markings, 
    mark= at position 0.6 with {\arrow{Straight Barb[ length=1mm, width=1mm]}}}
] 
  \tikzstyle{every node}=[circle, draw, fill=black,inner sep=0pt, minimum width=3pt]
  \node[align=left] (n2) at (9,0.5)  [label= right: $\phi_y$]{};
   \node[align=left]  (n1) at (9,0)  [label= right: ]{};
    \draw [postaction={decorate}](n2) -- (n3)  {} ;
    \draw [postaction={decorate}](n1) to[bend left=60] (n2)  {} ;
     \draw [postaction={decorate}](n1) to[bend right=60] (n2)  {} ;
      ; \node (n4) at (9.75,0.5) [label= right: $\phi_x$]{};
\end{tikzpicture}+\cdots\\
=&R_{\mathcal{T},gV}(\phi_x\phi_y).\end{split}
\end{equation}
\end{example}
  \begin{proof}[Proof to Lemma \ref{lemma: RT and operators}]
      By comparison of Lemmas \ref{lemma: diagrams operator expansion} and \ref{lemma:2008}, there is a bijection $f$ from the diagrams in the expansion of the operator product $(\hatphi_{x_1}^H)^{m_1}\ldots (\hatphi_{x_k}^H)^{m_k}$ to the diagrams in the expansion of $ R_{\mathcal{T},gV}(\phi_{x_1}^{m_1}\ldots \phi_{x_k}^{m_k})$, where the bijection deletes the external edges.

      Consider a diagram $D$ with $n$ internal vertices in the operator expansion. Then the value assigned by the operator Feynman rules to each directed edge in $D$ is the same as that assigned  by the algebraic Feynman rules to each directed edge in $f(D)$. Both schemes also assign the same coefficient $\frac{ig}{\hbar}c_z$ to internal vertices and the same summation over indices.

      In the algebraic approach, $f(D)$ is further assigned the constant
        $$\frac{\prod_{i=1}^n v(v-1)\cdots(v-v_i+1)\,\prod_{j=1}^k m_j(m_j-1)\cdots(m_j-w_j+1)}{\abs{\Aut(f(D))}}\,,$$
     where $v$ is the order of $V$, $v_i$ is the valency of the $i^\ith$ internal vertex in $f(D)$, and $w_j$ is the valency of the external vertex $x_j$ in $f(D)$. In the operator approach, $D$ is assigned the constant $\frac{H(D)}{\abs{\VP(D)}}$. That the two constant are equal follows from noting that the number of external edges at the $i^\ith$ vertex of $D$, $h_i$, is equal to $v-v_i$ and applying Lemmas \ref{lemma: aut} and \ref{lemma: H}.

     Finally, the $i^\ith$ internal (external) vertex in $f(D)$ is associated with a factor of $\phi^{v-v_i}$ ($\phi^{m_j-w_j}$, where $w_j$ denotes its valency). Applying the quantisation map $\QT$ turns the resulting functional into a time ordered product of field operators. This is the same product assigned to the external edges of $D$.
  \end{proof}
  
\begin{example} We illustrate the relationship between the operator expansion and the expansion of $R_{\mathcal{T},gV}$. Let $V=\frac{1}{3!}\sum_i\phi_i^3$. Then, 
\begin{multline*}
\QT R_{\mathcal{T},gV}(\phi_x) = \QT\bigg( \quad
 \begin{tikzpicture}[baseline=(current bounding box.center), decoration={markings,
    mark= at position 0.6 with {\arrow{Straight Barb[ length=1mm, width=1mm]}}}
] 
  \tikzstyle{every node}=[circle, draw, fill=black,inner sep=0pt, minimum width=3pt]
  \node (n3) at (9,0) [label= right: $\phi_x$]{};
\end{tikzpicture}
\ +\ 
\begin{tikzpicture}
[baseline=(current bounding box.center),decoration={markings, 
    mark= at position 0.8 with {\arrow{Straight Barb[ length=1mm, width=1mm]}}}
] 
  \tikzstyle{every node}=[circle, draw, fill=black,inner sep=0pt, minimum width=3pt]
  \node (n3) at (9,0.5) [label= right: $\phi_x$]{};
  \node(n2) at (9,0)  [label= right: ]{};
    \draw [postaction={decorate}](n2) -- (n3)  {} ;
\end{tikzpicture}
\quad + \;
\begin{tikzpicture}
[baseline=(current bounding box.center),decoration={markings, 
    mark= at position 0.6 with {\arrow{Straight Barb[ length=1mm, width=1mm]}}}
] 
  \tikzstyle{every node}=[circle, draw, fill=black,inner sep=0pt, minimum width=3pt]
  \node[baseline](n3) at (9,0.5) [label= right: $\phi_x$]{};
  \node (n2) at (9,0)  [label= right: ]{};
   \node (n1) at (9,-0.5)  [label= right: ]{};
    \draw [postaction={decorate}](n2) -- (n3)  {} ;
     \draw [postaction={decorate}](n1) -- (n2)  {} ;
\end{tikzpicture}
\quad +\ 
\begin{tikzpicture}
[baseline=(current bounding box.center),decoration={markings, 
    mark= at position 0.6 with {\arrow{Straight Barb[ length=1mm, width=1mm]}}}
] 
  \tikzstyle{every node}=[circle, draw, fill=black,inner sep=0pt, minimum width=3pt]
  \node[align=left] (n3) at (9,1) [label= right: $\phi_x$]{};
  \node[align=left] (n2) at (9,0.5)  [label= right: ]{};
   \node[align=left]  (n1) at (9,0)  [label= right: ]{};
    \draw [postaction={decorate}](n2) -- (n3)  {} ;
    \draw [postaction={decorate}](n1) to[bend left=60] (n2)  {} ;
     \draw [postaction={decorate}](n1) to[bend right=60] (n2)  {} ;
      ;
\end{tikzpicture}\quad +\; \begin{tikzpicture}
[baseline=(current bounding box.center),decoration={markings, 
   mark= at position 0.6 with {\arrow{Straight Barb[ length=1mm, width=1mm]}}}
] 
 \tikzstyle{every node}=[circle, draw, fill=black,inner sep=0pt, minimum width=3pt]
 \node (n3) at (9,0.7) [label= right: $\phi_x$]{};
 \node(n2) at (9,0)  [label= right: ]{};
  \node(n0) at (8.3,-0.7)  [label= left: ]{};
  \node(nx) at (9.8,-0.7)  [label= right: ]{};
   \draw [postaction={decorate}](n2) -- (n3)  {} ;
   \draw [postaction={decorate}](n0)  -- (n2)  {} ;
   \draw [postaction={decorate}](nx)  -- (n2)  {} ;
\end{tikzpicture}+\cdots\bigg)
\\
= \QT\bigg( \phi_x+\frac{g}{2} \, \Delta^R_{x,i_1}\phi_{i_1}^2 +\frac{g^2}{2}\Delta^R_{x,i_1}\Delta^R_{i_1,i_2}\big( \phi_{i_1}\phi_{i_2}^2-i\hbar\Delta^R_{i_1,i_2}\phi_{i_2}\big)+\frac{g^3}{8}\Delta^R_{x,i_1}\Delta^R_{i_1,i_2}\Delta^R_{i_1,i_3}\phi_{i_2}^2\phi_{i_3}^2+\cdots\bigg)
 \\
 = C\bigg[ \hatphi_x+\frac{g}{2} \, \Delta^R_{x,i_1}\hatphi_{i_1}^2 +\frac{g^2}{2}\Delta^R_{x,i_1}\Delta^R_{i_1,i_2}\big( \hatphi_{i_1}\hatphi_{i_2}^2-i\hbar\Delta^R_{i_1,i_2}\hatphi_{i_2}\big)+\frac{g^3}{8}\Delta^R_{x,i_1}\Delta^R_{i_1,i_2}\Delta^R_{i_1,i_3}\hatphi_{i_2}^2\hatphi_{i_3}^2+\cdots\bigg]
 \\
= \quad
 \begin{tikzpicture}[baseline=(current bounding box.center), decoration={markings,
    mark= at position 0.6 with {\arrow{Straight Barb[ length=1mm, width=1mm]}}}
] 
  \tikzstyle{every node}=[circle, draw, fill=black,inner sep=0pt, minimum width=3pt]
  \node (n3) at (9,0) [label= right: $\phi_x$]{};
  \draw [](9,-0.5) -- (n3)  {} ;
\end{tikzpicture}
\ +\ 
\begin{tikzpicture}
[baseline=(current bounding box.center),decoration={markings, 
    mark= at position 0.8 with {\arrow{Straight Barb[ length=1mm, width=1mm]}}}
] 
  \tikzstyle{every node}=[circle, draw, fill=black,inner sep=0pt, minimum width=3pt]
  \node (n3) at (9,0.5) [label= right: $\phi_x$]{};
  \node(n2) at (9,0)  [label= right: ]{};
    \draw [postaction={decorate}](n2) -- (n3)  {} ;
    \draw [](8.5,-0.5) -- (n2)  {} ;\draw [](9.5,-0.5) -- (n2)  {} ;
\end{tikzpicture}
\quad + \;
\begin{tikzpicture}
[baseline=(current bounding box.center),decoration={markings, 
    mark= at position 0.6 with {\arrow{Straight Barb[ length=1mm, width=1mm]}}}
] 
  \tikzstyle{every node}=[circle, draw, fill=black,inner sep=0pt, minimum width=3pt]
  \node[baseline](n3) at (9,0.5) [label= right: $\phi_x$]{};
  \node (n2) at (9,0)  [label= right: ]{};
   \node (n1) at (9,-0.5)  [label= right: ]{};
    \draw [postaction={decorate}](n2) -- (n3)  {} ;
     \draw [postaction={decorate}](n1) -- (n2)  {} ;\draw [](9.5,-0.5) -- (n2)  {} ;
     \draw [](8.5,-1) -- (n1)  {} ;\draw [](9.5,-1) -- (n1)  {} ;
\end{tikzpicture}
\quad +\ 
\begin{tikzpicture}
[baseline=(current bounding box.center),decoration={markings, 
    mark= at position 0.6 with {\arrow{Straight Barb[ length=1mm, width=1mm]}}}
] 
  \tikzstyle{every node}=[circle, draw, fill=black,inner sep=0pt, minimum width=3pt]
  \node[align=left] (n3) at (9,1) [label= right: $\phi_x$]{};
  \node[align=left] (n2) at (9,0.5)  [label= right: ]{};
   \node[align=left]  (n1) at (9,0)  [label= right: ]{};
    \draw [postaction={decorate}](n2) -- (n3)  {} ;
    \draw [postaction={decorate}](n1) to[bend left=60] (n2)  {} ;
     \draw [postaction={decorate}](n1) to[bend right=60] (n2)  {} ;
      ;\draw [](9,-0.5) -- (n1)  {} ;
\end{tikzpicture}\quad +\; \begin{tikzpicture}
[baseline=(current bounding box.center),decoration={markings, 
   mark= at position 0.6 with {\arrow{Straight Barb[ length=1mm, width=1mm]}}}
] 
 \tikzstyle{every node}=[circle, draw, fill=black,inner sep=0pt, minimum width=3pt]
 \node (n3) at (9,0.7) [label= right: $\phi_x$]{};
 \node(n2) at (9,0)  [label= right: ]{};
  \node(n0) at (8.3,-0.7)  [label= left: ]{};
  \node(nx) at (9.8,-0.7)  [label= right: ]{};
   \draw [postaction={decorate}](n2) -- (n3)  {} ;
   \draw [postaction={decorate}](n0)  -- (n2)  {} ;
   \draw [postaction={decorate}](nx)  -- (n2)  {} ;\draw [](8,-1) -- (n0)  {} ;\draw [](8.6,-1) -- (n0)  {} ;\draw [](10.1,-1) -- (nx)  {} ;\draw [](9.5,-1) -- (nx)  {} ;
\end{tikzpicture}+\cdots \quad
=\ \hatphi^H_x,
\end{multline*}
where the first and second lines of diagrams are evaluated via the algebraic and operator Feynman rules, respectively.
\end{example}
\subsection{$R_{H,gV}$ and in-in expectation values}
This section gives the diagrammatics of equation \eqref{proposition: in in}.

\begin{lemma}[Diagrammatic expansion of in-in correlators, from \cite{Albertini:2024srq}]
Given $x_i\not\prec x_j\in\calC$ for every $i<j$, some positive integers $m_1,\ldots, m_k$ and an interaction polynomial $V$,  the expectation value $_{\IN}\big\langle 0\big|(\phi_{x_1}^H)^{m_1}\ldots (\phi_{x_k}^H)^{m_k}\big|0\big\rangle_{\IN}$ is equal to the sum of half-edge graphs with no external edges that can be constructed from the graphs of Lemma \ref{lemma: diagrams operator expansion} by joining the external edges into undirected edges, where each graph is assigned a value via the operator Feynman rules.
\end{lemma}
\begin{lemma}[Diagrammatic expansion of $R_{H,gV}(\mathcal{T}(\phi_{x_1}^{m_1}\ldots \phi_{x_k}^{m_k}))(0)$]
$R_{H,gV}(\Tcal(\phi_{x_1}^{m_1}\ldots \phi_{x_k}^{m_k}))(0)$
is equal to the sum of operators $\vec{\gamma}(\phi_{x_1}^{m_1},\ldots, \phi_{x_k}^{m_k})(0)$ defined by the graphs that can be constructed from the graphs of Lemma \ref{lemma:2008} by adding undirected edges such that the external vertices have valencies $m_1,\cdots,m_k$ and the internal vertices have valency equal to the order of $V$, where each graph is assigned a value via the algebraic Feynman rules.
\end{lemma}

\begin{proof}
    It follows from equation \eqref{relation moller} and Corollary \ref{lemma T graph} that $R_{H,gV}(\Tcal(\phi_{x_1}^{m_1}\ldots \phi_{x_k}^{m_k}))$
is equal to the sum of operators $\vec{\gamma}(\phi_{x_1}^{m_1},\ldots, \phi_{x_k}^{m_k})$ defined by the graphs that can be constructed from the graphs of Lemma \ref{lemma:2008} by adding undirected edges in all possible ways. When setting $\phi=0$, only the graphs with valencies as in the Lemma statement are non-vanishing.
\end{proof}

\begin{corollary}\label{cor: in-in diagrams}
The diagrammatic expansions for expressions ${}_\IN\langle0|(\hatphi^H_{x_1})^{m_1}\cdots(\hatphi^H_{x_k})^{m_k}|0\rangle_\IN$ and $R_{H,gV}(\Tcal(\phi_{x_1}^{m_1}\cdots\phi_{x_k}^{m_k}))(0)$ in the two lemmas above coincide. In particular, this gives a diagrammatic proof of Proposition~\ref{proposition: in in ordered}.
\end{corollary}
\begin{proof}
Apply $\langle0|\cdot|0\rangle$ to Lemma~\ref{lemma: RT and operators} and use $\langle0|\QT(F)|0\rangle=(\Tcal F)(0)$ (Proposition~\ref{proposition: T Tbar} with $l=0$) together with \eqref{relation moller}. At the level of graphs, the bijection $f$ of the proof of Lemma~\ref{lemma: RT and operators} extends to the graphs obtained by joining the external edges into undirected edges, respectively by adding undirected edges. Both sets of rules assign $\hbar\Delta^F$ to undirected edges, and the combinatorial constants agree by Lemmas~\ref{lemma: H} and~\ref{lemma: aut}.
\end{proof}

\section{Path integral expressions}\label{sec: path integral}
The following theorem identifies the independently defined Gaussian history integral with the algebraic star product.

\begin{theorem}\label{thm:path integral} For $F,G\in\CC[\calC]$, with the polynomial integral defined in Section~\ref{sec:pathint}, equation~\eqref{eq: moment formula},
    \begin{equation}\begin{split}
    \overline{\mathcal{T}}G\starH\mathcal{T}F(0)=\int_{\mathbb{R}^{2N}}d^N\phi^-\,d^N \phi^+ \ G(\phi^-) \  D_0(\phi^+,\phi^-)\ F(\phi^+).
\end{split}
\end{equation}
\end{theorem}

\begin{proof}
Put $u=(\phi^-,\phi^+)$. Evaluating the Gaussian Fourier derivatives in the polynomial-moment formula \eqref{eq: moment formula} gives
\[
\int d^{2N}u\,D_0(u) G(\phi^-)F(\phi^+)
=\left[e^{\frac\hbar2\partial_u^T\Sigma_0\partial_u}G(\phi^-)F(\phi^+)\right]_{u=0}.
\]
Only finitely many derivative terms act nontrivially on $G(\phi^-)F(\phi^+)$. The two diagonal blocks of $\Sigma_0$ give $\overline\Tcal$ acting on $G$ and $\Tcal$ acting on $F$; the two off-diagonal blocks give $\exp(\hbar\partial_-^TW\partial_+)$, which is precisely the contraction defining $\starH$. This proves the stated identity without using any operator representation of the integral.
\end{proof}

\begin{corollary}
   Given a real local $V\in\CC[\calC]$ and $F\in\CC[\calC]$, with the following equality understood coefficientwise in $g$,
   \begin{align}
R_{H, gV}(\mathcal{T}F)(0)=\int_{\mathbb{R}^{2N}}d^N\phi^-\,d^N \phi^+ \ \mathscr{S}^{-1}(\phi^-) \  D_0(\phi^+,\phi^-)\ \mathscr{S}(\phi^+)F(\phi^+).
\end{align}
\end{corollary}
\begin{proof}
Set $G=\mathscr{S}^{-1}$ and replace $F$ by $\mathscr S F$ in Theorem~\ref{thm:path integral}, then use the unitarity argument \eqref{eq: unitarity} to eliminate the $\overline{\Tcal}$ on the LHS and compare with definition \ref{def moller} of the M\o ller operator.
\end{proof}
\begin{corollary}\label{cor: in-in path integral}
For real local $V\in\CC[\calC]$ and $F,G\in\CC[\calC]$, the in-in formula \eqref{eq: df master equation1} holds coefficientwise in $g$.
\end{corollary}
\begin{proof}
By Proposition~\ref{proposition: T Tbar interacting} (extended by linearity to polynomials), the left-hand side of \eqref{eq: df master equation1} equals $\overline\Tcal(\mathscr S^{-1}G)\starH\Tcal(\mathscr SF)(0)$. Apply Theorem~\ref{thm:path integral} with $G$ replaced by $\mathscr S^{-1}G$ and $F$ by $\mathscr SF$, coefficient-wise in $g$.
\end{proof}
This proof does not use the operator representation \eqref{eq:dcf0}, so is an independent cross-check of the previous results. Similarly, one can prove

\begin{corollary}\label{cor: reduction}
For real local $V\in\CC[\calC]$, $F\in\CC[\calC]$ and $k,l\geq0$, the reduction formula \eqref{eq: reduction} holds coefficientwise in $g$.
\end{corollary}
\begin{proof}
By Proposition~\ref{prop: matrix element}, extended by linearity from monomials to polynomials $F$, the left-hand side of \eqref{eq: reduction} equals $\sum_{M\in\mathfrak M_{k,l}}c_M\hbar^{d_M/2}[\Tcal\mathcal D_M(\mathscr SF)](0)$. Theorem~\ref{thm:path integral} with $G=1$ gives $(\Tcal Q)(0)=\int d^N\phi^-d^N\phi^+\,D_0(\phi^+,\phi^-)\,Q(\phi^+)$ for every polynomial $Q$, since the $\phi^+\phi^+$ block of $\Sigma_0$ is $\Delta^F$. With $Q=\mathcal D_M(\mathscr SF)$, coefficientwise in $g$,
\[
[\Tcal\mathcal D_M(\mathscr SF)](0)
=\int d^N\phi^-d^N\phi^+\,
D_0(\phi^+,\phi^-)\,\mathcal D_M^+\bigl(\mathscr SF\bigr)(\phi^+).
\]
Now $\mathcal D_M^+$ is a constant-coefficient differential operator of order $d_M$ acting on $\phi^+$. Define the moments of $\partial_u^\beta D_0$ by the same cut-off limit as in Section~\ref{sec:pathint}. Then
\[
\langle\partial_u^\beta D_0,P\rho_R\rangle=(-1)^{|\beta|}\langle D_0,\partial_u^\beta(P\rho_R)\rangle .
\]
In every term in which a derivative falls on $\rho_R$, that derivative produces a factor $R^{-1}$ times a rescaled Schwartz function. The pairing of $D_0$ with such a term stays bounded as $R\to\infty$, so these terms vanish in the limit. Hence $\int(\partial_u^\beta D_0)P=(-1)^{|\beta|}\int D_0\,\partial_u^\beta P$. Moving the $d_M$ derivatives onto $D_0$ produces the sign $(-1)^{d_M}$ and gives \eqref{eq: reduction}.
\end{proof}

\section{Conclusions}\label{sec: conclusions}
We have developed a perturbative description of interacting scalar quantum field theory on a finite causal set, relating its algebraic, operator and histories formulations. The quantisation maps and their associated products provide a common framework for expressing Wick contractions, causal ordering and anti-causal ordering. Within this framework, we obtain expressions for in-in and in-out correlation functions and relate particle transition amplitudes to field insertions.

\section*{Acknowledgments}
Stav Zalel would like to thank Fay Dowker for helpful discussions and the University of York for the Mathematics Fellowship Bursary that financed a visit where some of the research took place. Kasia Rejzner would like to thank ICMS in Edinburgh and the organisers of the workshop ``Women in Mathematical Physics 3,'' where she had numerous important discussions with Fay Dowker, Emma Albertini and Sumati Surya. 

\appendix
\section{Diagrammatic expansion of operators}\label{appendix: sec diagrams}
We prove Lemma \ref{lemma: diagrams operator expansion}. We will need the following definition and lemmas from \cite{Albertini:2024srq,Transtrum:2005}.
\begin{definition}[The map $\Upsilon$]\label{Upsilon definition appendix} Given an ordered list of natural numbers, $\textbf{a}=a_1a_2\ldots a_n$, we define,
\begin{equation}
  \Upsilon(\textbf{a})= \begin{cases}
      \frac{1}{l_1!\ldots l_k!} & a_1\geq a_2 \geq \ldots \geq a_n\\
      0 & \text{otherwise,}\\
    \end{cases}    
\end{equation}
where $k$ denotes the number of strings of equal signs in $a_1\geq a_2\geq \ldots\geq  a_n$ and $l_i$ is the length of the $i^\ith$ string of equal signs.
\end{definition} For example, $\Upsilon(8,7,5,2,1)=1, \Upsilon(7,7,5,4,4,4)=\frac{1}{2!1!3!}$ and $\Upsilon(7,5,7,4,4,4)=0$.

\begin{lemma}\cite[Corollary A.5]{Albertini:2024srq}\label{cor operator product expansion}
Given a finite causal set $\calC$ and $k$ integers satisfying $|\calC|\geq x_1>x_2\ldots >x_k\geq 1$, let $\mathcal{O}^H(x_i)$ denote a local Heisenberg picture operator living at $x_i$. Then the operator product $\mathcal{O}^H(x_1)\ldots \mathcal{O}^H(x_k)$ can be expressed as the following sum of nested commutators, 
    \begin{equation}\label{eq: cor operator product expansion}
    \begin{split}
  &\mathcal{O}^H(x_1)\ldots \mathcal{O}^H(x_k)\\        &=\sum_{n=0}^{\infty} \bigg(\frac{-ig}{\hbar}\bigg)^n\sum_{n_1+\cdots +n_{k}=n}\ \ \sum_{z_1,\ldots, z_{n_1}=x_2}^{x_1-1}\ \ \sum_{z_{n_1+1},\ldots, z_{n_2}=x_3}^{x_2-1}\cdots \ \ \sum_{z_{n_{1}+\cdots+ n_{k-1}+1},\ldots, z_{n}=1}^{x_{k}-1}\\
&\bigg[\hspace{-1mm} \cdots \bigg[\mathcal{O}(x_1),V(z_1)\bigg]\hspace{-1mm} \cdots\hspace{-0.8mm} ,V(z_{n_1})\bigg]\mathcal{O}(x_2),V(z_{n_1+1})\bigg]\hspace{-1mm} \cdots\hspace{-0.8mm}  \mathcal{O}(x_k),V(z_{n_1+\cdots+ n_{k-1}+1})\bigg]\hspace{-1mm} \cdots\hspace{-0.8mm}  ,V(z_{n})\bigg]\Upsilon(\textbf{z}),
    \end{split}
    \end{equation}
    where on the RHS, $\mathcal{O}$ and $V$ are polynomials in the free field operators, and the $n=0$ term is understood to be $\mathcal{O}(x_1)\ldots \mathcal{O}(x_k)$.
\end{lemma}
\begin{lemma}\cite[Eq. (23)]{Transtrum:2005}\label{lemma commutator}

Consider an algebra of operators $P_1,P_2,...$ satisfying the commutation relations $[P_i,P_j]=c_{ij}$, and let $A\equiv A(P_1,\ldots,P_{n-1})$ and $B\equiv B(P_n)$ be polynomials in $P_1,\ldots, P_{n-1}$ and $P_n$, respectively. Then,
\begin{equation}\begin{split} \label{commutator_identity}
\big[A,B\big]=& - \sum_{k=1}^{\infty} \ \ \sum_{k_1+\cdots+k_{n-1}=k} \ \ \bigg(\prod_{l=1}^{n-1} \frac{(-c_{l,n})^{k_l}}{k_l!}\bigg)\ \partial_{P_1}^{k_1}\cdots\partial_{P_{n-1}}^{k_{n-1}}A \ \partial_{P_n}^k B.
\end{split}
\end{equation}
\end{lemma}

\begin{corollary}[Diagrammatic representation of the commutator]
\label{cor: commutator diagrams}
\

The commutator $\big[A,B\big]$ can be represented as the sum of all half-edge graphs $G$ with,
\begin{itemize}
    \item $n$ vertices labelled $1,\ldots, n$,
    \item for $l=1,\ldots,n-1$, the valency of the vertex $l$ is the order of $P_l$ in $A$; the valency of the vertex $n$ is the order of $P_n$ in $B$,
    \item all edges are directed from $n$ to $l$ for some $l=1,\ldots, n-1$,
    \item there is at least one directed edge.
\end{itemize}

\hspace{-5mm}Each graph is interpreted as follows:
\begin{itemize}
\item each directed edge $l\leftarrow n$ gives a factor of $-c_{ln}$,
\item for $l$ with $l=1,\dots,n$, each external edge at vertex $l$ contributes a factor of $P_l$; the product of these operators is ordered as it was ordered in $A$ and then with $P_n$ on the far right,
\item multiply by $H(G)$ (as defined in \ref{def H}),
\item multiply by an overall factor of $-1$.
\end{itemize}
\end{corollary}

\begin{corollary}\label{cor: caus ord prod}
  The operator product $\mathcal{O}^H(x_1)\ldots \mathcal{O}^H(x_k)$ of Lemma \ref{cor operator product expansion} can be expressed as a causally ordered expansion in free fields.
\end{corollary}
\begin{proof}
    Lemma \ref{lemma commutator} can be applied recursively to express the nested commutator in \eqref{eq: cor operator product expansion} as a sum of monomials in free field operators, where in each monomial the fields are ordered from left to right by the causal set point on which they live, as: $$x_1,\ z_1,z_2,\ldots,z_{n_1},\ x_2,\ z_{n_1+1},\ldots, z_{n_1+\ldots+n_k},\ x_k,\ z_{n_1+\ldots+n_k},\ldots, z_n.$$ The limit of the sums and the factor of $\Upsilon(\mathbf{z})$ in \eqref{eq: cor operator product expansion} ensure that this ordering is causal.
\end{proof}

\begin{proof}[Proof of Lemma \ref{lemma: diagrams operator expansion}]
     The upshot of corollary \ref{cor: caus ord prod} is that --- since each monomial is causally ordered, and since we have $\Delta^R=0$ on the diagonal --- the factors of the field commutators $c_{x,y}=i\hbar\Delta_{x,y}$ that appear in the expansion can be replaced by factors of the retarded propagator, $i\hbar\Delta^R_{x,y}$. This in turn enables us to change the limits of the sums in \eqref{eq: cor operator product expansion} to $z_i=1$ to $x_1-1$ for all $i=1,\ldots,n$. The diagrammatic representation of the commutator given in corollary \ref{cor: commutator diagrams} is then applied recursively to obtain a diagrammatic representation for the nested commutator in \eqref{eq: cor operator product expansion} (where arrows are now interpreted as retarded propagators). Finally, the factor of $\Upsilon(\mathbf{z})$ is absorbed into the diagrams by the method of \cite[Eq. (4.2-4.5)]{Albertini:2024srq}, yielding the diagrammatic rules for the product expansion.
\end{proof}
\printbibliography

\end{document}